\documentclass[10 point,preprint,linenumbers]{aastex631}
\usepackage{graphicx,amsmath}

\usepackage{float}
\usepackage{hyperref}
\usepackage{url}
\usepackage{tikz}
\usepackage{soul}
\setstcolor{blue}
\usepackage{rotating}
\usepackage{xcolor}
\usepackage{lineno}
\received{XXXX}
\revised{XXXX}
\accepted{XXXX}

\newcommand{\RomanNumeralCaps}[1] {\MakeUppercase{\romannumeral #1}}

\shorttitle{MHD evolution of a jet}

\begin{document}

\title{Exploring the magnetic and thermal evolution of a coronal jet}
\correspondingauthor{Sushree S. Nayak, Samrat Sen}
\author[0000-0002-4241-627X]{Sushree S. Nayak}
\affil{Center for Space Plasma \& Aeronomic Research,
The University of Alabama in Huntsville,
Huntsville, Alabama 35899, USA}
\email{corresponding email ids: s.sangeetanayak93@gmail.com, samratseniitmadras@gmail.com}

\author[0000-0003-1546-381X]{Samrat Sen}
\affil{Instituto de Astrof\'{i}sica de Canarias, 38205 La Laguna, Tenerife, Spain}

\author[0000-0001-9035-3245]{Arpit Kumar Shrivastav}
\affil{Aryabhatta Research Institute of Observational Sciences, Nainital, India-263002}
\affil{Joint Astronomy Programme and Department of Physics, Indian Institute of Science, Bangalore 560012, India}

\author[0000-0003-4522-5070]{R. Bhattacharyya}
\affil{Udaipur Solar Observatory, Physical Research Laboratory,
Dewali, Bari Road, Udaipur 313001, India}

\author[0000-0002-4454-147X]{P.S. Athiray}
\affil{Center for Space Plasma \& Aeronomic Research,
The University of Alabama in Huntsville,
Huntsville, Alabama 35899, USA}
\affil{NASA Marshall Space Flight Center, ST13, Huntsville, AL 35812, USA}

\begin{abstract}
Coronal jets are the captivating eruptions which are often found in the solar atmosphere, and primarily formed due to magnetic reconnection. Despite their short-lived nature and lower energy compared to many other eruptive events, e.g. flares and coronal mass ejections, they play an important role in heating the corona and accelerating charged particles. However, their generation in the ambience of non-standard flare regime is not fully understood, and warrant a deeper investigation, in terms of their onset, growth, eruption processes, and thermodynamic evolution. Toward this goal, this paper reports the results of a data-constrained three-dimensional (3D) magnetohydrodynamics (MHD) simulation of an eruptive jet; initialized with a Non-Force-Free-Field (NFFF) extrapolation and carried out in the spirit of Implicit Large Eddy Simulation (ILES). The simulation focuses on the magnetic and dynamical properties of the jet during its onset, and eruption phases, that occurred on February 5, 2015 in an active region NOAA AR12280, associated with a seemingly three-ribbon structure. In order to correlate its thermal evolution with computed energetics, the simulation results are compared with differential emission measurement (DEM) analysis in the vicinity of the jet. Importantly, this combined approach provides an insight to the  onset of reconnection in transients in terms of emission and the corresponding electric current profiles from MHD evolutions. The presented study captures the intricate topological dynamics, finds a close correspondence between the magnetic and thermal evolution in and around the jet location. Overall, it enriches the understanding of the thermal evolution due to MHD processes, which is one of the broader aspects to reveal the coronal heating problem.

\end{abstract}


\section{Introduction}
\label{intro}
Solar coronal jets are one of the fascinating eruptions from the solar atmosphere alongside flares and coronal mass ejections (CMEs). They are collimated plasma flows which are morphologically inverted-Y in shape, short-lived, and energetically 4-5 orders less than flares \citep{2013ApJ...776...16P, 2016SSRv..201....1R}. The study of jets is essential as they play an important role in heating the corona, and accelerating the solar wind \citep{1983ApJ...272..329B,2007Sci...318.1591S, 2019Sci...366..890S}.

The generation mechanism behind these jets is primarily ascribed to magnetic reconnection which involves the reconfiguration of magnetic field topology while releasing stored magnetic energy into Ohmic heating, kinetic energy and accelerating the charged particles from the site of reconnection. Fully or partially open magnetic field lines near the coronal holes or at the boundary of active regions are observed as the favorable sites for ejecting jets while the triggering may be either through flux cancellation \citep{2008A&A...481L..57C, 2011ApJ...738L..20H, 2014PASJ...66S..12Y, 2016ApJ...822L..23P, 2017ApJ...844..131P, 2018ApJ...853..189P, 2019:panesar, 2019ApJ...882...16M, 2017ApJ...844...28S} or flux emergence \citep{1992PASJ...44L.173S, 2007Sci...318.1591S, 2008ApJ...673L.211M, 2021NatCo..12.6621M}. In one of the important studies by \citet{2015Natur.523..437S}, a plausible scenario is reported for the eruption of coronal jets which are now also extensively applicable to many similar bursts of different energetic scales. According to these studies, a pre-existing minifilament escapes the canopy of surrounding open and closed field lines while reconnecting both at that site as well as at its own footpoints. The plasma material ejects out in that collimated channel and is dubbed as a spire carrying a swirling motion. These type of jets are popularly known as ``blowout jets'' \citep{2010ApJ...720..757M,2013ApJ...769..134M}. In addition to that, jets are often associated with an initial bright point, also known as a jet bright point (JBP) \citep{2012ApJ...745..164S, 2014PASJ...66S..12Y, 2015Natur.523..437S, 2016ApJ...822L..23P, 2017ApJ...844...28S} before achieving the fully grown structure.

Flare ribbons are inherent to solar eruptions. The well-known standard flare model a.k.a.~the CSHKP model \citep{1964NASSP..50..451C, 1966Natur.211..695S, 1974SoPh...34..323H, 1976SoPh...50...85K} pictures a solar eruption as a consequence of release of a magnetic flux rope above the polarity inversion line (PIL) from the solar atmosphere in two dimensions (2D). When the flux rope overcomes the tension force of overarching magnetic loops, that creates a current sheet below ensuing reconnection and producing two ribbons at the footpoints of the reconnected field lines. While this is the case for standard flares irrespective of their eruptive nature, there exists non-standard flares which produces different kind of ribbons namely circular/quasi-circular flare ribbons \citep{2009ApJ...700..559M, 2012ApJ...760..101W, prasad+2018apj, 2023A&A...674A.154M}, X-shape ribbons \citep{2016ApJ...823L..13L, 2016NatSR...634021L}, and three-ribbon types \citep{2014ApJ...781L..23W}.

These non-standard type ribbons or flares do not fit the argument of the standard flare model in explaining the flaring or eruptive process. In fact, in the state-of-art scenario, it is not fully understood about its onset, growth, and eruption processes, and merits attention to study. Besides, if they are associated with jet or CME, it is again challenging to comprehend the eruption process, formation of current sheets, or the reconnection affair in broad to capture in three-dimensional (3D) models. But, several attempts with an ensemble of multiple observations, and theoretical/numerical modeling are being made to understand aforementioned criticalities involved in the whole eruption process in these types of atypical events. Especially, when the responsible magnetic configuration in these events is analyzed to detect a potential reconnection triggering location, several potential configurations are reported as an initiation mechanism. For example, \citet{2009ApJ...700..559M} has observed the formation of current sheets ``due to stressing of spine'' of the magnetic null point (ideally, $|\mathbf{B}| =0$) skeleton in their simulation which was associated with a quasi-circular flare ribbon. Likewise, \citet{prasad+2018apj} has explained the role of magnetic null point and quasi-separatrix layers (QSLs; locations bearing a sharp change in magnetic connectivities; \citep{1995JGR...10023443P, 1996A&A...308..643D}) in triggering a circular flare ribbon in the active region AR 12192 via magnetohydrodynamics (MHD) simulation initialized with a non-force-free-field (NFFF) extrapolation. Similarly, \citet{2014ApJ...781L..23W} discussed a three-ribbon flare produced due to reconnection along the coronal null line. In a different scenario of an X-shaped ribbon in \citet{2016NatSR...634021L}, a hyperbolic flux tube (HFT; intersection of two QSLs; \citep{2002JGRA..107.1164T, 2007ApJ...660..863T, 2012ApJ...750...15S}) was found in the non-linear-force-free-field (NLFFF) constructed magnetic topology where the dissipation of current sheets leads to ribbons.

With the above backdrop, in this paper, we focus on understanding the onset of an eruptive jet followed by an atypical flare ribbon type or majorly the reconnection process while analyzing the magnetic and thermal evolutions with a side-by-side comparison with observational dynamics. 
The novelty of this approach lies in its envisaged effectiveness in emphasizing the triggering mechanism, the dissipation of magnetic energy, and decay of current and its implications; on the emission near the jet region.

We study a particular event of jet eruption which occurred on February 5, 2015, in AR12280 as reported in \citet{2023A&A...672A..15J}. The above study was mainly concentrated to investigate the oscillation of a filament that was hit by an erupting jet. However, the exploration of magnetic field topology and the other magnetic properties (e.g. field line twist, field strength, ohmic heating, etc.) are not explored due to the constraint of spectroscopic imaging. The role of magnetic properties becomes significant at low plasma-$\beta$ in solar corona, and hence it warrants an importance to study the magnetic properties in and around the jet during the eruption using MHD simulations. It also motivates us to enrich the general understanding of the onset process of transients.

The above discussions emphasize the role of magnetic topology and the sites of reconnection in 3D for any transients. However, direct and routine measurements of the coronal magnetic field are still unavailable. Therefore, magnetic field extrapolation has become an alternative to obtain the coronal magnetic field topology using available photospheric magnetic fields. Myriad numerical attempts have been made to perform realistic modeling of coronal magnetic field such as potential source surface models \citep{1969SoPh....9..131A, 1969SoPh....6..442S, 2006ApJ...653.1510R}, several force-free models \citep{1999A&A...350.1051A, 2004SoPh..222..247W, 2006SoPh..235..201I, 2007SoPh..245..251W, 2007MmSAI..78..126R, 2011ApJ...727..101J, 2012ApJ...759...85J, 2012LRSP....9....5W, 2014A&ARv..22...78W}, non-force-free models \citep{1993SoPh..143...41Z, 1994SoPh..151...91Z, 1999SoPh..186..123G}, magneto-frictional approaches \citep{2012ApJ...757..147C, 2016ApJ...828...82G}, and magnetohydrostatic models \citep{2003SoPh..214..287W, 2007A&A...475..701W, 2008A&A...481..827R} which includes both static and time-dependent solutions. For further details on features of different coronal models, we direct the readers to \citet{2013SoPh..288..481R, 2017SSRv..210..249W} and references therein. In this work, we have adapted the non-force-free-field extrapolation model devised by \citet{2008ApJ...679..848H,2010JASTP..72..219H} to construct the coronal magnetic field topology. Then, we have initiated a data-constrained MHD simulation with the NFFF extrapolated magnetic field using the EULAG (Eulerian/Semi-Langrangian) MHD model \citep{smolarkiewicz&charbonneau2013jcoph} to track the eruption process.  In our study, we have given effort to deliver a collective analysis of the events using magnetic field extrapolation, MHD modeling, and comparison of the modeled parameters with the derived observables  near the jet, and ribbons.

The EULAG-MHD code is based on an incompressible regime, and not capable of estimating thermodynamic evolution directly. This motivates us to use Differential Emission Measurement (DEM) technique to estimate the thermal evolution of the jet region and its surroundings from observational data. This method is extensively used to evaluate the thermodynamic information for various solar observations in general. Also, particularly for the coronal jet studies, it has been used to explore the thermodynamic evolution of the spire, footprints, and source regions \citep{2016Mulay,2017Mulay, 2019Mulay, 2023Yang, 2023Zhang}.

The remainder of the paper is as follows. Section \ref{event} outlines the event of an active region jet eruption that occurred. In Section \ref{dem}, we have reported the results obtained from DEM analysis of the event. Section \ref{ext} explains the rationale behind the NFFF extrapolation model and discusses the modeled extrapolated field of AR12280. Section \ref{simu} details the EULAG-MHD model, required set-ups, and simulated results. Lastly, Section \ref{conc} summarizes the key findings from the work, highlights the novelty of the study,  and concludes how this work can be useful in a broader aspect to understand the solar atmosphere in future studies.

\section{The Jet observed on 2015 February 5 in the active region AR12280}
\label{event}

\begin{figure}[h!]
\begin{center}

\vspace{1cm}
\includegraphics[width=\textwidth]{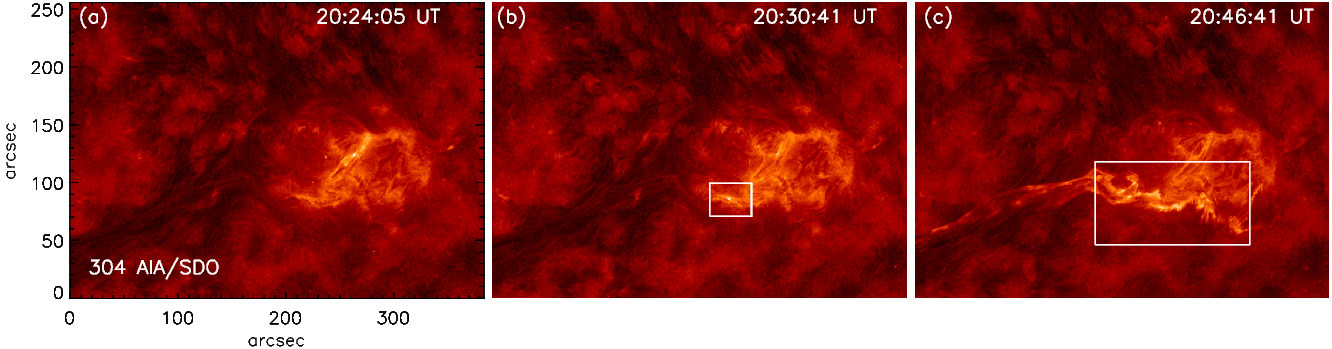}
\includegraphics[width=.5\textwidth]{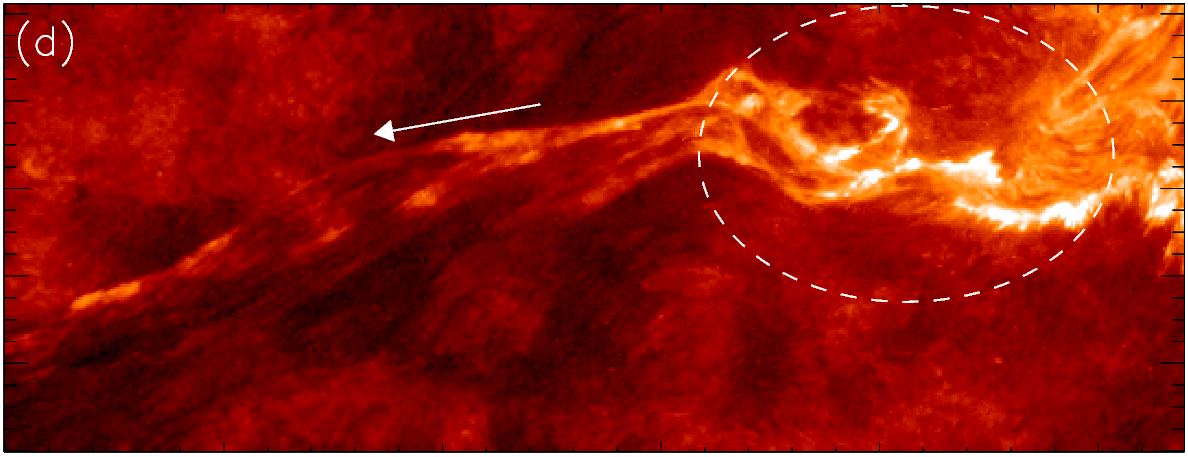}
\caption{History of the jet formation in 304 {\AA} channel of AIA/SDO. Panel (a) is the initial stage before the onset of the jet whereas panel (b) marks the onset of the jet near the JBP location enclosed in a white box. The fully grown structure of the jet with distinct base and spire is plotted in panel (c).  The jet base region and the three-ribbon structure are marked by a bigger white box. In panel (d) we have provided the zoomed in view of the jet at $\approx$ 20:46 UT marking its base and spire regions with a circle and an arrow in white respectively. An animation showing the jet formation with the same field of view as the still images is provided. The animation starts at 20:24:07 UT and ends at 21:03:19 UT. The real-time duration is 13 seconds.} 
\label{jetobs}
    \end{center}
\end{figure}

\begin{figure}\centering
\includegraphics[width=.45\textwidth]{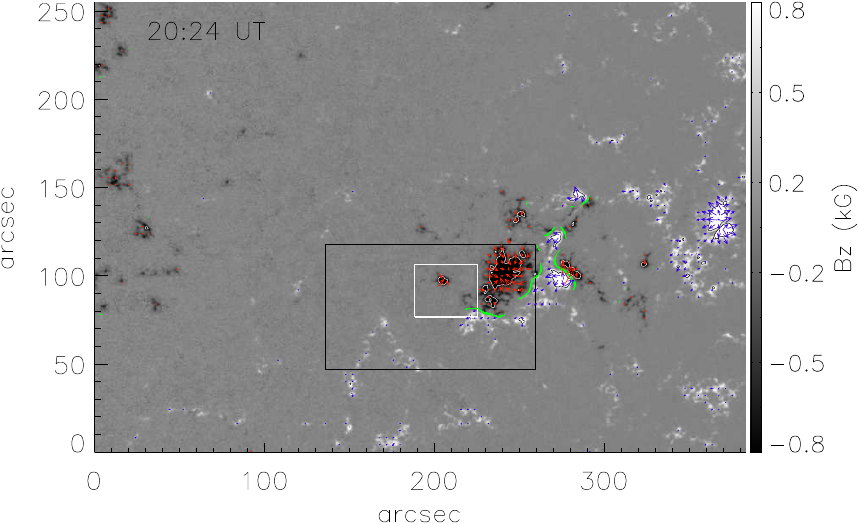} 
\includegraphics[width=.45\textwidth]{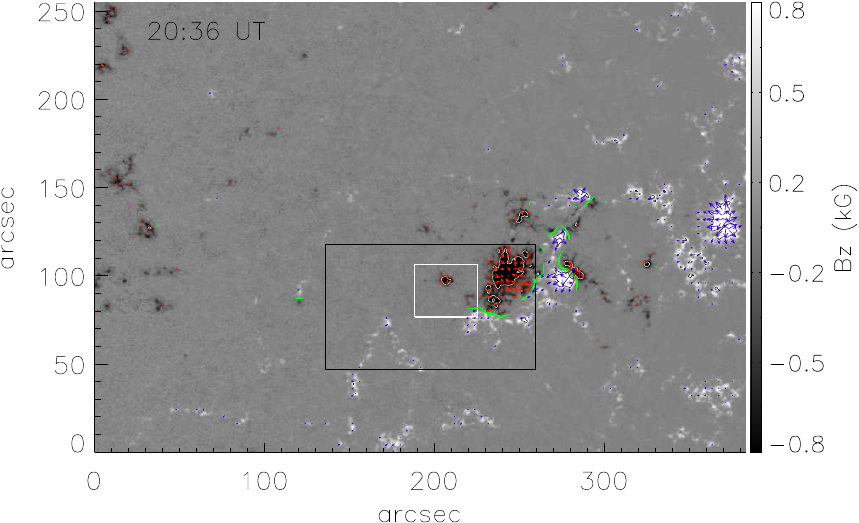} 
\caption{Left panel: magnetogram of AR12280 at 20:24 UT with the same field of view as Figure \ref{jetobs}. Notably, magnetogram at this timestamp is used for extrapolation further. Right panel: magnetogram at 20:36 UT, close to onset phase of the jet, as marked in panel (b) of Figure \ref{jetobs}. The blue and red arrows represent the vector plots of positive and negative transverse components over the $B_{z}$ component in gray ranging from (-800G, 800G). The green contours show the polarity inversion lines. The black box denotes the jet base region and the white box covers the bright point region.} 
\label{hmibz}
\end{figure}

We have studied the jet in the active region AR12280 on 2015 February 5 peaked at $\approx$ 20:42 UT. In Figure \ref{jetobs}, we have plotted the jet covering the whole active region in the field of view (FOV) in 304 {\AA} pass-band of Atmospheric Imaging Assembly (AIA; \citep{Lemen2012}) onboard Solar Dynamics Observatory (SDO; \citep{pesnell+2012soph}). Panel (a) corresponds to a time instance where no sign of jet activity is detected. Around $\approx$ 20:30 UT, we spotted an appearance of a bright point in the east side of the AR as marked by the white box in panel (b), which is referred to as JBP. This is a signature of the onset of the jet, which is developed afterward as shown in panel (c) around $\approx$ 20:46 UT. The jet base region is highlighted by the white box there. The legible three-ribbon structure can also be noticed. Panel (d) depicts the zoomed view of the jet highlighting its base region by the dotted white circle and the direction of the spire by the white arrow. An animation is provided showing the evolution of the jet in 304 channel. The animation starts at 20:24:07 UT and ends at 21:03:19 UT. We have plotted the $B_{z}$ component of AR12280 at 20:24 UT obtained from Helioseismic Magnetic Imager (HMI; \citep{2012SoPh..275..229S}) in Figure \ref{hmibz}. Similarly, we have marked the jet base region in the black box and the bright point location in the white box respectively on the $B_{z}$ map. The blue and red arrows indicate the direction of the transverse components on the $B_{z}$ map plotted in gray scale. The green color contours are the location of polarity inversion lines (PILs) in the active region. The PILs are plotted using the routine developed by \citet{2018SoPh..293...16S}. According to the routine, firstly, all of the pixels in the magnetogram are convolved with a gaussian kernel. Then, centering each pixel of $B_{z}$ component, five consecutive pixels both in horizontal and vertical directions are scanned and maximum and minimum values are compared in both the directions of those five pixels. If they are found to be of opposite sign and magnitude of these two values exceeds the noise level of 60 Gauss (by trial and error method) on either side of the two arrays, the pixel in between them is located as PIL. In the right panel, we have plotted the magnetogram closer in time to the onset of the jet i.e. 20:36 UT as marked in Figure \ref{jetobs}.

\section{DEM analysis near the Jet region}
\label{dem}
To understand the thermodynamic changes in the region of interests (ROIs) (see Figure \ref{jetobs}: bright point region (white box in panel (b), jet base region (bigger white box in panel (c)), we determine differential emission measure (DEM) distributions using six optically thin emission\sout{s} in AIA pass-bands. These channels have contribution from ionized states of iron with wavelengths 94  \AA\ (Fe \RomanNumeralCaps{10}, Fe \RomanNumeralCaps{18}), 131 \AA\ (Fe \RomanNumeralCaps{8}, Fe \RomanNumeralCaps{21}), 171 \AA\ (Fe \RomanNumeralCaps{9}), 193 \AA\ (Fe \RomanNumeralCaps{12}, Fe \RomanNumeralCaps{24}), 211 \AA\ (Fe \RomanNumeralCaps{14}), and 335 \AA\ (Fe \RomanNumeralCaps{16}). Their temperature response function peaks at log T[\emph{K}] = (6.05, 6.85), (5.6, 7.05), 5.85, (6.2, 7.25), 6.3, and 6.45 respectively \citep{2010ODwyer, 2012Boerner}. AIA channels are sensitive to a wide range of plasma temperatures. It is also established that the bi-modal temperature response of the hot AIA channels (94 \AA\ , 131 \AA\ , 193 \AA\ ) can introduce systematic overestimation of emission measure at high temperatures \citep{2024ApJ...961..181A}. We perform DEM analysis using 16 temperature bins from 5.4 $\leq$ log T $\leq$ 7.0, with $\delta$ log T = 0.1.

We have used the latest version of sparse inversion code \citep{2015Cheung, 2018Su} to obtain DEM, which accounts for the amount of plasma emission in particular temperature intervals integrated along our line of sight.

\begin{equation}
	DEM (T) =  n_{e}^{2} \frac{dl}{dT} ,
\end{equation}
where \emph{l} is the path length along the line of sight and $n_{e}$ denotes the electron density.

\begin{figure}[h]
\centering
\includegraphics[width=1\textwidth]{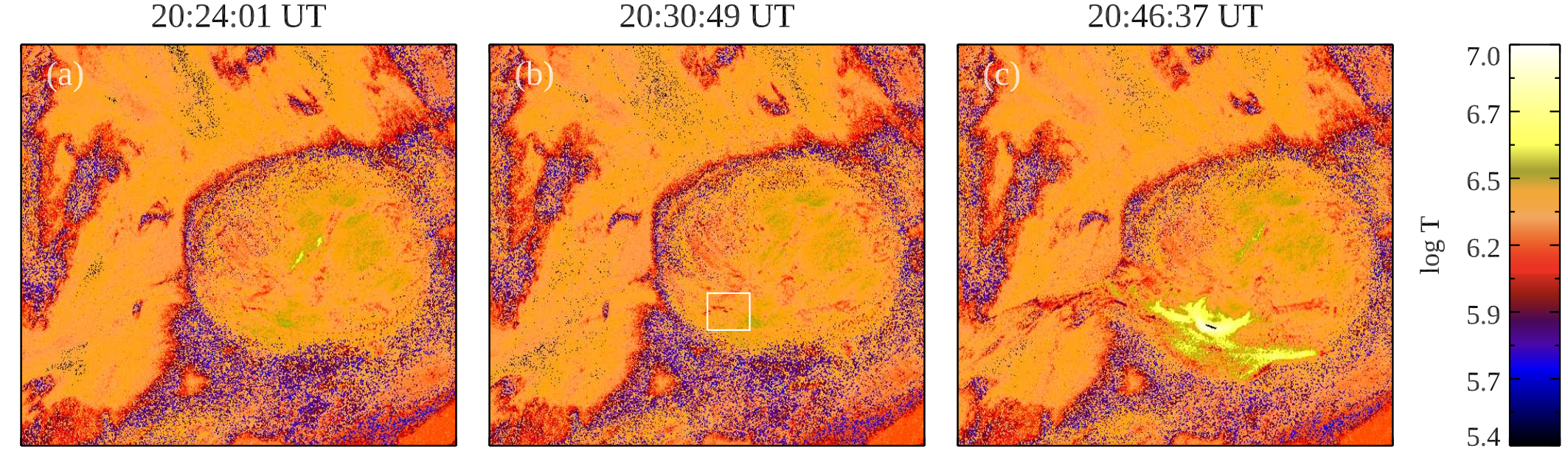} 
\caption{Evolution of EM weighted log(T) over the active region till the growth of the jet region. Like Figure \ref{jetobs}, the panel (a) denotes the state before the onset of the transients. Panel (b) marks the bright point region in the white box and panel (c) shows the hot loops in the flux rope eruption region which also comprises of the base part of the jet.} 
\label{temp_maps}
\end{figure}

\begin{figure}[h]
\centering
\includegraphics[width=1\textwidth]{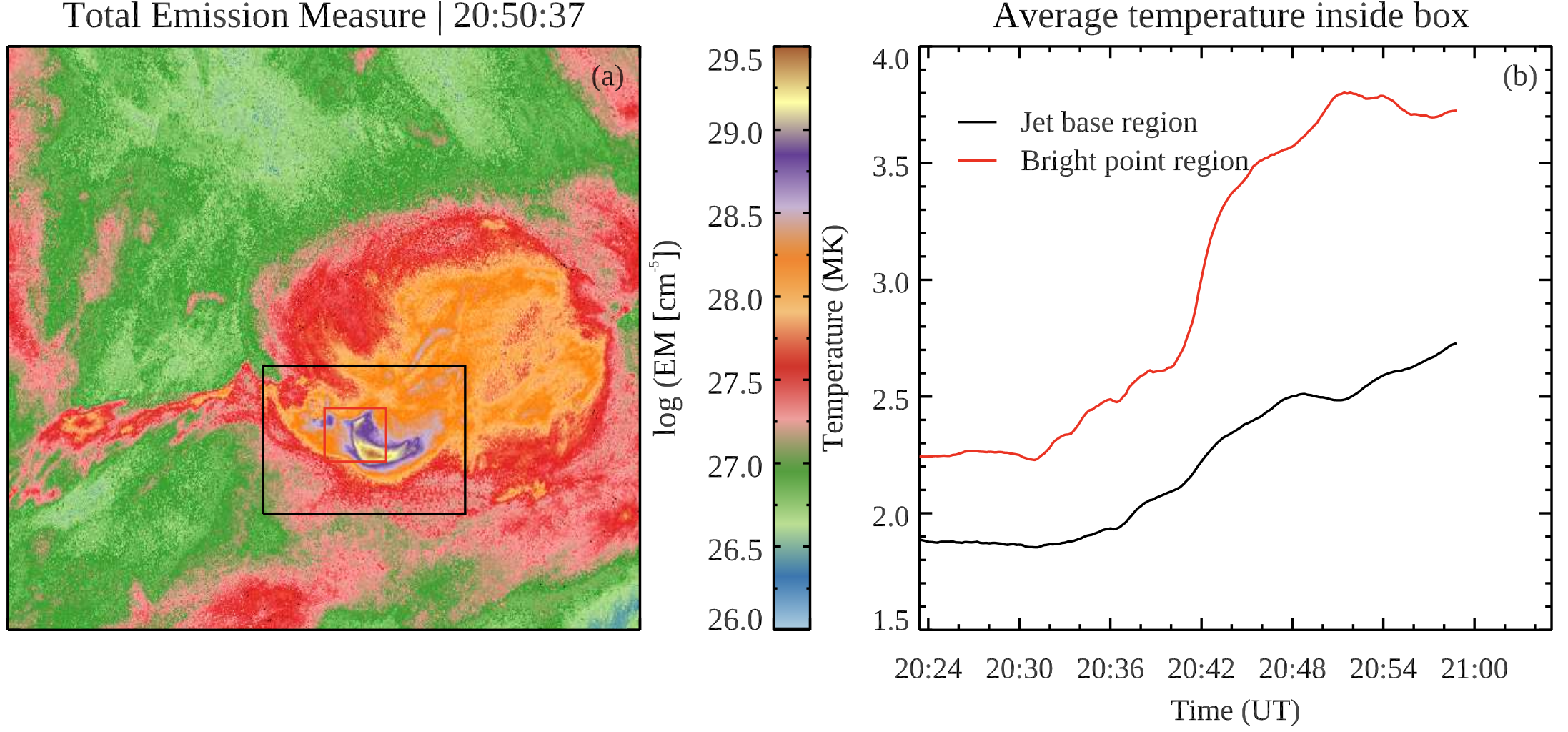} 
\caption{Time evolution of EM weighted temperature near the jet base and bright point regions from DEM analysis. The left panel displays the total emission measure map within the Regions of Interest at 20:50:37 UT. The regions under black and red boxes are utilized to determine the average temperature in the right panel. Evident there is a sharp increase in the maximum temperature near the bright point region than the jet base.} 
\label{demtemp}
\end{figure}

The total emission measure can be calculated from $DEM(T)$,
\begin{equation}
	EM = \int DEM(T) \times dT ~~.
\end{equation}
 The $DEM(T)$ in every temperature bin can be utilized to obtain emission measure weighted temperature as,
\begin{equation}
	T_{EM} = \frac{\int T\ DEM(T)\ dT}{\int DEM(T)\ dT}~~.
\end{equation}
 We have plotted the evolution of the EM weighted temperature extracted through DEM analysis in Figure \ref{temp_maps} for the whole active region. Panel (a) is the temperature distribution on the whole region. The bright point location is marked in panel (b). Later in panel (c), it can be seen that the hotter regions are exhibiting near the location of filament eruption. For further inspection, in Figure \ref{demtemp}, we plotted the emission map in logarithmic scale in panel (a) highlighting the ROIs in black and red boxes.
 In panel (b), we have plotted the average EM weighted temperature inside those two boxes till the decay phase of transients.  
Important is the higher and a sharp ascent in the peak of the temperature near the bright point region than the jet base.

\section{Initial magnetic configuration of AR12280}
\label{ext}

\subsection{The Non-Force-Free-Field Extrapolation Model}
\label{nfff}
To investigate the onset of the jet eruption and the formation of multiple ribbon structures, we first obtained the initial magnetic field topology of the AR12280 using the Non-Force-Free-Field (NFFF) extrapolation model \citep{2008ApJ...679..848H,2010JASTP..72..219H}. The NFFF model is theorized based on the minimum dissipation rate principle (MDR) where the idea uses total dissipation rate as a minimizer keeping the generalized helicity constant for a two-fluid description of plasma \citep{2004PhPl...11.5615B, 2007SoPh..240...63B}. The relaxed state carries a non-zero Lorentz force which is used to drive the plasma in the MHD simulation later in our study. The extrapolation solves an inhomogeneous double-curl
Beltrami equation for the magnetic field $\mathbf{B}$ (\citet{2004PhPl...11.5615B, 2007SoPh..240...63B}, and references therein),

\begin{equation}
 	\nabla \times \nabla \times \mathbf{B} + a_{1} \nabla \times \mathbf{B} + b_{1} \mathbf{B} = \nabla \psi,
\end{equation}
\noindent
where $a_{1}$ and $b_{1}$ are constants. The solenoidality of $\mathbf{B}$ enforces
the scalar function $\psi$ to obey Laplace's equation. The modified
vector $\mathbf{B'} = \mathbf{B} - \nabla \psi$ satisfies the corresponding homogeneous
equation, which represents a two-fluid MHD steady state \citep{1998PhRvL..81.4863M} having a solution \citep{2006GeoRL..3315106H, 2008ApJ...679..848H}
\begin{equation}
 	\mathbf{B'} = \sum_{i=1,2} \mathbf{B}_{i},
\end{equation}
\noindent where each $\mathbf{B}$ is a linear force-free field satisfying
\begin{equation}
 	\nabla \times \mathbf{B}_{i} = \alpha_{i} \mathbf{B}_{i},
\end{equation}
\noindent in usual notations. The two sets of constants are related by
$a_{1} = - (\alpha_{1}+\alpha_{2})$ and $b_{1} = \alpha_{1}\alpha_{2}$. The magnetic field is then
\begin{equation}
 	\mathbf{B} = \mathbf{B}_{3} + \sum_{i=1,2} \mathbf{B}_{i},
\end{equation}
\noindent where $\mathbf{B}_{3} = \nabla \psi$ is a potential field.
We apply the technique described in \citep{2010JASTP..72..219H}. Briefly, an optimal pair of $\alpha$ is
computed for $\mathbf{B}_{3} = 0$ by minimizing the average normalized
deviation of the magnetogram transverse field $B_{t}$ from its
extrapolated value $b_{t}$, given by
\begin{equation}
	E_{n} =  \left( \sum_{i=1}^M |\mathbf{B}_{t,i} - \mathbf{b}_{t,i}| \times |\mathbf{B}_{t,i}|\right) \bigg/\left(\sum_{i=1}^M |\mathbf{B}_{t,i}|^2 \right),
\end{equation}
\noindent where $M = N \times N$ is the total number of grid points on the
transverse plane. Additional minimization of $E_{n}$ is done by
using $\mathbf{B}_{3} = \nabla \psi$ as a corrector field for the obtained pair of $\alpha$'s.
Noting a superposition of potential fields results in potential
field only, the above procedure is iteratively repeated until 
$E_{n}$ plotted against the number of iterations asymptotically
saturates to a minimum. Also
important is to recognize that $\alpha_{1}$ and $\alpha_{2}$ , alone or combined,
are not the only parameters to determine field line twist \citep{2018ApJ...869...69M}. With the
twist $\tau$ being related to the field-aligned current density
\begin{equation}
	\tau = \frac{\mathbf{J} \cdot \mathbf{B}}{|\mathbf{B}|^2} = \frac{(\alpha_{1} \mathbf{B}_{1} + \alpha_{2} \mathbf{B}_{2}) \cdot (\mathbf{B}_{1} + \mathbf{B}_{2} + \mathbf{B}_{3})}{|\mathbf{B}|^2},
\end{equation}
additional to any modifications in $\alpha_{1}$ and $\alpha_{2}$, $\tau$ can also vary
because of changes in the component fields—  including the
potential field $\mathbf{B_{3}}$, an advantage of the NFFF extrapolations.
We note that the values of $\alpha_{1}$ and $\alpha_{2}$ must be bounded above to
ensure a monotonous decay of $\mathbf{B}$ with height \citep{1971SoPh...19...72N}. If required, extra twist can be accommodated in the
extrapolated $\mathbf{B}$ by varying $\mathbf{B}_{1}$, $\mathbf{B}_{2}$ and $\mathbf{B}_{3}$ to better match the
magnetogram while not exceeding the maximal $\alpha_{i}$'s.

The justification of using the NFFF model lies in the following analysis. As per \citet{2001SoPh..203...71G} on his study of plasma beta, $\beta = \frac{p}{p_{mag}}$ (where, $p$ and $p_{mag}$ represent the plasma and magnetic pressure respectively) variation over the active regions, the force-free approximations on the photosphere does not hold true. There in Figure-3, only the mid corona, a sandwiched region between chromosphere and the upper corona/solar wind accelerated region is shown to have a $\beta$ value of less than unity. One of the important arguments for the non-force-free-field approximation lies in the use of magnetic field. \citet{1995ApJ...439..474M} have shown that the magnetic field is not force-free at the photosphere while becomes the same at a height around 400 km above the photosphere. \citet{2002ApJ...568..422M} have studied the nature of the photopshere in 12 flare producing active regions and found it not to be much deviant from force-free nature. However this may not hold true always if there are twisted structures. In \citet{2012ApJ...744...65T}, they also found similar arguments but in umbral or inner penumbral regions using high spatial resolution magnetograms from the Solar Optical Telescope/Spectro-Polarimeter on
board Hinode. Besides, for the available force-free extrapolation models based on optimization and MHD relaxation approaches, there can remain a finite residual Lorentz force during the minimization to an equilibrium state. As a remedy to this issue, a preprocessing technique is adapted in the algorithm where the photospheric magnetic field is adjusted. Whereas, the NFFF algorithm operates on observed magnetograms without any apriori preprocessing keeping the original nature of force at the photosphere. Again, the NFFF model does not claim to provide the whole active region to be non-force free. Furthermore, one rationale can be provided in terms of the strength of the Lorentz force, which depends on the magnetic field strength as well as the associated current density (which depends on the spatial gradient of the magnetic field components). Also, the orientation between $\bf{B}$ and $\bf{J}$ contribute to strength of the Lorentz force. Therefore, we may not always conclude the force-free or non-force-freeness of a magnetic field configuration based only on the magnetic field strength.

\subsection{The NFFF Extrapolated Magnetic Field Topology}
\begin{figure}
\centering
\includegraphics[width=.6\textwidth]{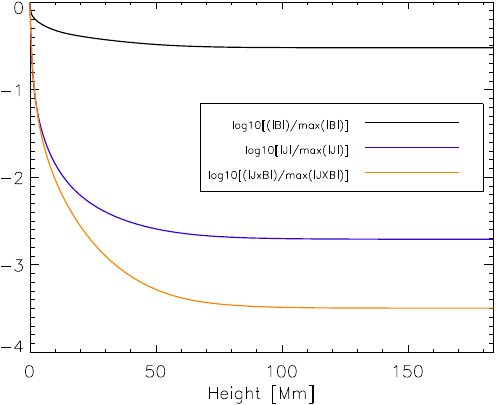}
\caption{Variation of the magnitude of average magnetic field ($|\bf{B}|$), current density ($|\bf{J}|$), and Lorentz force ($|\bf{J}\times \bf{B}|$) over height in logarithmic scale. }
\label{line_field}
\end{figure}

To extrapolate the magnetic field over AR12280, we have used the \texttt{hmi.B\_720}s magnetogram series of HMI. The data in this series consists of magnetic field strength (B), azimuth, and inclination in every 720 seconds with a full disk FOV.  We have reproduced an HMI SHARP-like active region patch from B, azimuth, and inclination using the \texttt{bvec2cea.pro} available in the \texttt{Solarsoft} package \citep{2012:solarsoft} with a slight modification in the coordinate system to it. The objective is to cover a field of view that encloses the base part of the jet sufficiently for the extrapolation while maintaining a flux balance over the cutout. The computational box has 768 uniform grid points or $\approx$ 276 Mm of physical extent in the x-direction and 512 grid points or $\approx$ 184 Mm in both y- and z- directions. The $E_{n}$ has a value of $\approx .36$  which amounts to $\approx$ 36\% error in the reconstruction of the transverse field whereas the $B_{Z}$ is the same observed magnetic field as stated in the Section \ref{nfff}. The Figure \ref{line_field} depicts the variation of magnetic field ($|\bf{B}|$), current density ($|\bf{J}|$), and Lorentz force ($|\bf{J}\times \bf{B}|$) over the geometrical height normalized to their maximum value in logarithmic scale over the whole computational domain. We have also calculated $\beta$ inside the jet base region for a pixel having the maximum total B of value $\approx$2790 G and found to be $\approx .44$ at the photospheric level and for a field strength of $\approx 500$ G, we found it to be 14. In conclusion, the active region does have a range of $\beta$ and may not hold the force-freeness condition everywhere. Importantly, the structures in our extrapolated field lies over the PILs where a sharp gradient of the magnetic field connectivity is observed.

\begin{figure}
\centering
\includegraphics[width=\textwidth]{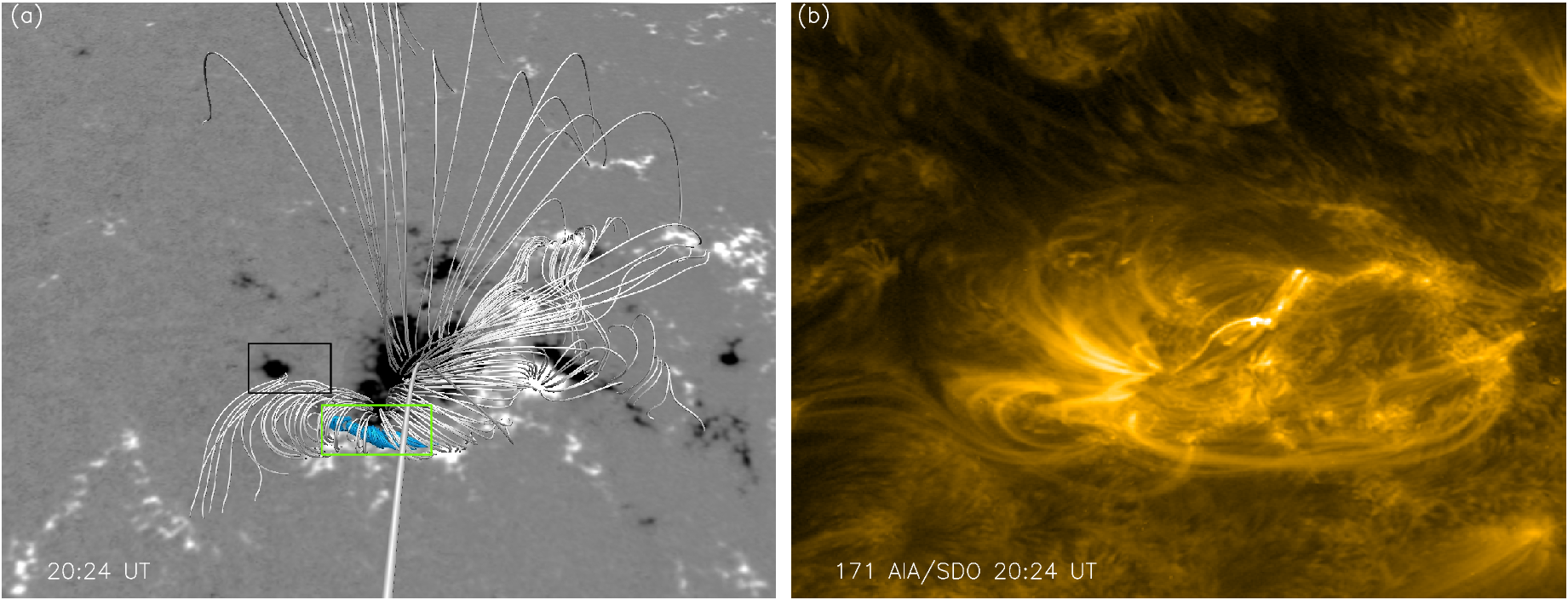}
\includegraphics[width=\textwidth]{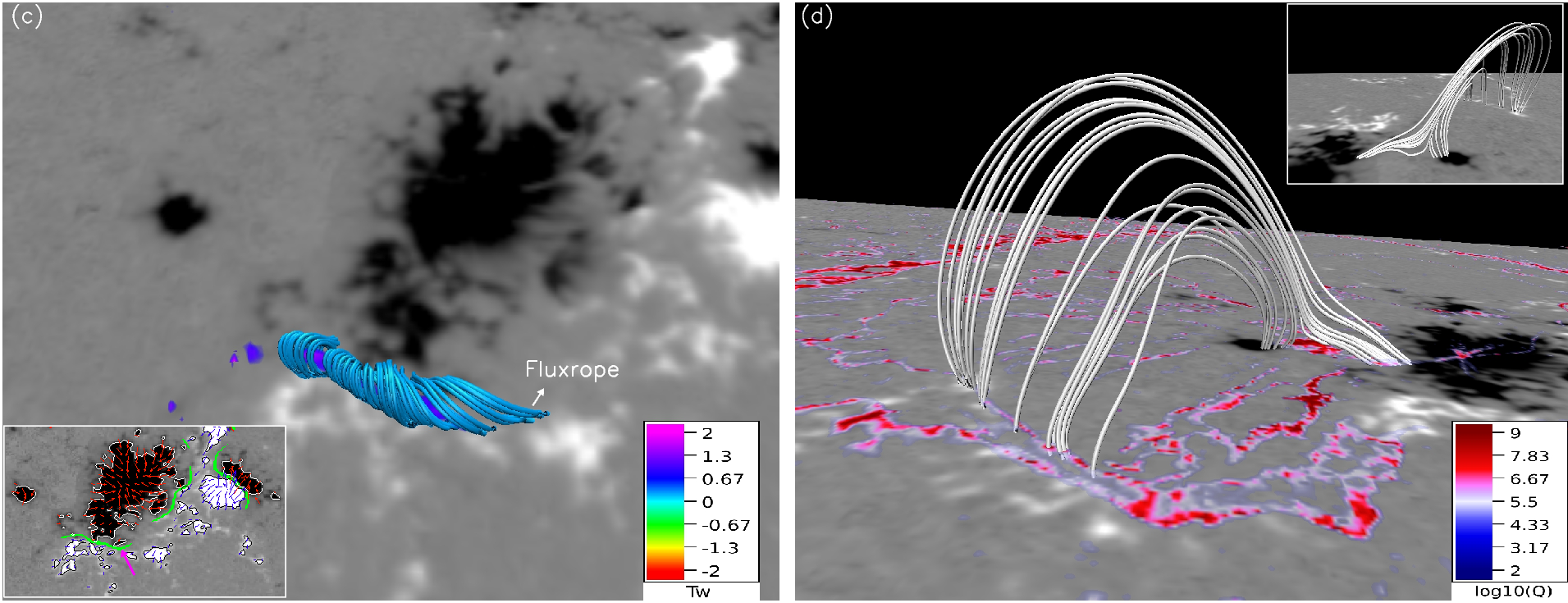}
\caption{(a) Field lines plotted over the active region magnetogram with $B_{z}$ component on the background in gray scale; (1) cyan color twisted field lines over the base of the jet region, and (2) random field lines (white) over the active region. Panel (b) shows the loops in the 171 {\AA} channel near the active region at the same instance. The green color box in panel (a) represents the field lines constituting the flux rope whose zoomed in version is plotted in panel (c) with Direct Volume Rendering (DVR) of the twist ($T_{w}$) parameter indicating a maximum value along the rope exceeding unity. The inset here is a cutout of the magnetogram shown in Figure 2 left panel. The green lines inside the inset mark the locations of the PILs, and the magenta arrow points toward the region where the flux rope is detected. The bald patch is highlighted in a black box in panel (a) and the field lines surrounding it are plotted in panel (d) with a distribution of squashing factor or $Q$-map on the background. Regions with high $Q$-value are found along the footpoints of the bald patch. We can also see the bald patch from a flipped side in the inset of panel (d).}
\label{init}
\end{figure}

In Figure \ref{init}, we have plotted the magnetic field lines using the Visualisation and Analysis Platform for Ocean, Atmosphere, and Solar Researchers (VAPOR) software \citep{2019Atmos..10..488L}. The field lines are plotted employing a highly accurate Field-line integration function of VAPOR,  which relies on the adaptive line integration with a fourth-order Runge–Kutta scheme and tri-linear interpolation of field values over cells in the grid; described in Section 2.3.3 of \citet{clyne-2007}. In the extrapolated field, we find a flux rope near the jet base region, plotted in cyan color in the panel (a)  of Figure \ref{init}, also highlighted in the chartreuse color box. To confirm the structure, we have plotted a direct volume rendering of the twist parameter ($T_{w}$) near the flux rope. For this purpose, we have used the code by \citet{liu-2016}, also available at (http://staff.ustc.edu.cn/~rliu/qfactor.html). The  twist value is measurement of two infintesimally close field lines winding about each other \citep{2006JPhA...39.8321B} and can be casted in the form according to \citet{liu-2016};
\begin{eqnarray}
	T_{w} = \int_{L} {\frac{\mathbf{J} \cdot \mathbf{B}}{|\mathbf{B}|^2}} dl.
\end{eqnarray}
\noindent This flux rope here is a realization of the mini-filament found in case of jets as reported in \citet{2015Natur.523..437S}. We will discuss further dynamics in the MHD evolution in Section \ref{disc}. Near the flux rope, we find the maximum twist to be of $\approx 1.33$, shown in panel (c). The flux rope is found to be lying close to the bottom boundary. We have provided an inset there highlighting the location of the PIL, indicated by a magenta arrow, where the flux rope is detected. We find a bald patch type topology to the east of the flux rope, highlighted in the black box in panel (a). In panel (b), we have presented the 171 {\AA} channel for the same field of view and at same time instance as panel (a) to mark the resemblance of the overall topology of the active region. We then compute the squashing factor or the $Q$-value near these regions utilizing the routine of \citet{liu-2016}. We find the presence of QSLs in the vicinity of flux rope and bald patch. The footpoints of the bald patch are found to trace the high $Q$-value regions which is evident from the panel (d). The inset in panel (d) shows a flipped side view of the orientation of field lines in the bald patch.

\section{Framework of the EULAG-MHD model and the Numerical Set-up for MHD Evolution}
\label{simu}
\label{setup}
To track the evolution of the field lines near the jet base region and its effect on the eruption process, we have performed an MHD simulation using the EULAG-MHD model. The model solves the incompressible Navier-Stokes MHD equations under the assumption of thermal homogeneity (or thermally inactive) and perfect electrical conductivity \citep{sanjay2016}. The governing equations are as follows:

\begin{eqnarray}
\label{stokes}
& & \frac{\partial{\mathbf{v}}}{\partial t} 
+ \left({\mathbf{v}}\cdot\nabla \right) {\mathbf{ v}} =-\nabla p
+\left(\nabla\times{\mathbf{B}}\right) \times{\mathbf{B}}+\frac{\tau_\textrm{a}}{\tau_\nu}\nabla^2{\mathbf{v}},\\  
\label{incompress1}
& & \nabla\cdot{\mathbf{v}}=0, \\
\label{induction}
& & \frac{\partial{\mathbf{B}}}{\partial t}=\nabla\times({\mathbf{v}}\times{\mathbf{B}}), \\
\label{solenoid}
 & & \nabla\cdot{\mathbf{B}}=0, 
\label{e:mhd}
\end{eqnarray}
\noindent written in usual notations. The variables in the MHD equations are normalized as follows 
\begin{equation}
\label{norm}
{\mathbf{B}}\longrightarrow \frac{{\mathbf{B}}}{B_0},\quad{\mathbf{v}}\longrightarrow \frac{\mathbf{v}}{v_\textrm{a}},\quad
 L \longrightarrow \frac{L}{L_0},\quad t \longrightarrow \frac{t}{\tau_\textrm{a}},\quad
 p  \longrightarrow \frac{p}{\rho {v_\textrm{a}}^2}. 
\end{equation}
\noindent The constants $B_0$ and $L_0$ are generally arbitrary, but they can be fixed using the average magnetic field strength and size of the system. Here, $v_\textrm{a} \equiv B_0/\sqrt{4\pi\rho_0}$ is the Alfv\'{e}n speed and $\rho_0$ is the constant mass density. The constants $\tau_\textrm{a}$ and $\tau_\nu$ represent the Alfv\'{e}nic transit time ($\tau_\textrm{a}=L_0/v_a$) and viscous dissipation time scale ($\tau_\nu= L_0^2/\nu$), respectively, with $\nu$ being the kinematic viscosity. Utilizing the discretized incompressibility constraint, the pressure perturbation, denoted by $p$,  satisfies an elliptic boundary-value problem on the discrete integral form of the momentum equation (Equation \ref{stokes}) \citep{bhattacharyya+2010phpl} and the references within.

Here we discuss only essential features of the EULAG-MHD and refer readers to \citet{smolarkiewicz&charbonneau2013jcoph} and references therein. The model is based on the spatio-temporally second-order accurate non-oscillatory forward-in-time multidimensional positive definite advection transport algorithm MPDATA \citep{smolarkiewicz2006ijnmf}.  
Importantly, MPDATA has proven the dissipative property which, intermittently and adaptively, regularizes the under-resolved scales by simulating magnetic reconnections
and mimicking the action of explicit subgrid-scale turbulence models \citep{2006JTurb...7...15M} in the spirit of
Implicit Large Eddy Simulations (ILES)
\citep{grinstein2007book} scheme. Arguably, the residual numerical dissipation is then negligible
everywhere but at the sites of MRs. Moreover, this dissipation being intermittent in time and space, a quantification of it is meaningful only in
the spectral space where, analogous to the eddy viscosity of
explicit subgrid-scale models for turbulent flows, it only acts on the shortest modes admissible on the grid; in particular, in the vicinity of steep gradients in
simulated fields. Such ILESs conducted with the model have already been successfully utilized to simulate reconnections to understand their role in the coronal dynamics \citep{prasad+2017apj,prasad+2018apj,2019ApJ...875...10N, 2021PhPl...28b4502N, 2022FrASS...939061K, 2022ApJ...925..197B}. In this work, the presented computations continue to rely on the effectiveness of ILES in regularizing the under-resolved scales by the commencement of magnetic reconnections.

The initial magnetic field is supplemented from the NFFF extrapolation and the initial velocity field is set to ${\mathbf{v}}=0$. The lateral boundaries ($x$, and $y$) are kept open aiming that the net magnetic flux should be conserved. While at the bottom boundary, the $z$-components of ${\mathbf{B}}$ and  ${\mathbf{v}}$ are chosen to be fixed to their initial values (also termed as line-tied boundary condition) as the flux change during the transient activities is found to be minimum. The top boundary follows the same condition as bottom only exception of it not being fixed throughout the evolution. Also, except the bottom boundary, all variables are calculated by linearly extrapolating from the immediate neighborhood cell values. Notably, the field and the corresponding Lorentz force values at such height become extremely small compared to the counterparts at the lower boundaries \citep{prasad+2017apj, prasad+2018apj, 2019ApJ...875...10N, 2021PhPl...28b4502N}. As stated earlier in Section \ref{ext}, the simulation is initially driven by the non-zero Lorentz force associated with the extrapolated magnetic field, and the flow is  primarily generated by it. The resulting flow is however made incompressible following the Equation \ref{incompress1}, an assumption also adapted by \citet{1991ApJ...383..420D, 2005A&A...444..961A}. Since our focus is to understand the onset of the jet through the topological changes, the assumption seems to be justifiable in the tenuous coronal medium. One reminder is that the density is set to unity and being constant over both space and time. The computational box extension is of the same size as that of the extrapolation, but with $\approx 2$ times reduced scaling of $\delta x$. The spatial unit length $\delta x$ is .0052 and the time step $\delta t$ is set to $1\times 10^{-3}$ while satisfying the CFL condition \citep{1967IBMJ...11..215C}. The dimensionless coefficient or the kinematic viscosity $\tau_a$/$\tau_\nu \approx 2 \times 10^{-4}$ in the simulation is roughly $\approx$15 times larger than its coronal value \citep{prasad+2018apj}. To note, the parameter $\tau_a$/$\tau_\nu$ is controlled by the spatial resolution and the time step  whilst satisfying the von Neumann stability criteria \citep{1947PCPS...43...50C}. The larger $\tau_\textrm{a}$/$\tau_\nu$, however, only expedites the evolution without an effect on the corresponding changes in the magnetic topology \citep{prasad+2018apj, 2022FrASS...939061K}. The total simulation time is 5400$\delta t$. To compare with the observational time, we multiplied the total simulation time by 15. Then, it corresponds to a $\approx 40.5$ minutes of the observational period.

\section{Results and Discussions}
\label{disc}
\subsection{Dynamics in the field line topologies near all ROIs}
\begin{figure}
\centering
\includegraphics[width=.5\textwidth]{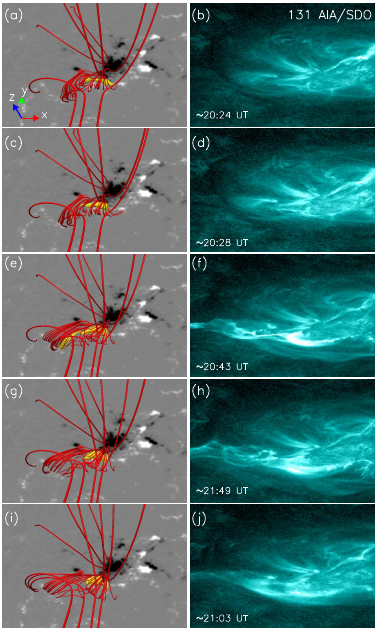}
\caption{Side-by-side comparison of overall magnetic field lines evolution with 131 {\AA} channel of AIA/SDO for the total period of simulation, $t \in \{20:24, \sim 21:03\}$ UT. The opening of the flux rope on the east end of it is visible in panel (e). Panels (g) and (i) show the scenario after the release of the plasma materials. An animation is provided to visualize the close correspondence between the simulated field lines and the observed dynamics in 131 channels. The animation begins at 20:24:08 UT and ends at 21:03:20 UT. The real-time duration of the video is 13 s.}
\label{evolfield}
\end{figure}

\begin{figure}
\centering
\includegraphics[width=.8\textwidth]{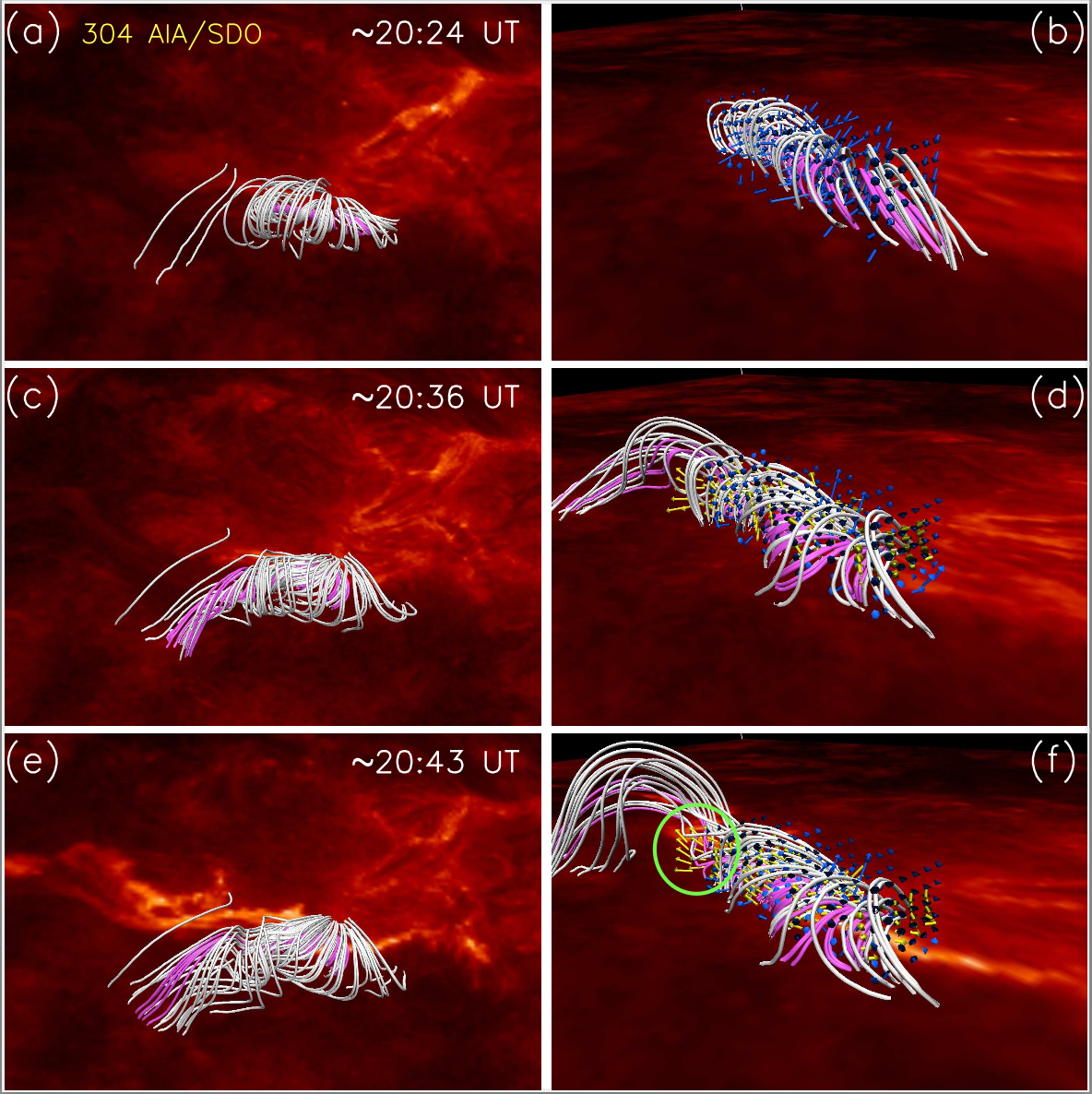}
\caption{Left column, Panels (a), (c) and (e): evolution of the flux rope (pink color) till the formation of flare ribbons overplotted with 304 {\AA} channel on the background.  Right column: Panels (b), (d), and (f) highlight the directions of the Lorentz force (blue color) and plasma flow (yellow color) surrounding the flux rope and the arcade (white color) above it. The orientation of the Lorentz force in panels (b) and (c) indicates the direction of the plasma erupted due to the untwisting of the flux rope. Important is also the rise of the surrounding magnetic arcade during the eruption of the filament here. The green color circle highlights kinks in the white and pink color field lines, as an apparent indication of internal reconnection near the footpoints of the flux rope. An animation is provided for the zoomed-in view of flux rope evolution surrounding the Lorentz force and plasma flow as depicted in the still images from panels (b), (d), and (f) but with $B_{z}$ on the background. The animation starts at 20:24 UT and ends at 20:42 UT. The real-time duration of the animation is 6 seconds.
} 
\label{ribbon}
\end{figure}

We plotted the overall evolution of magnetic topologies near the periphery of the jet and ribbons in Figure \ref{evolfield} vis-\`{a}-vis with the observed dynamics extracted from 131 {\AA} channel of AIA for the whole simulation period. The animation associated with Figure \ref{evolfield} is available online, that belongs from $t = 20:24$ to $21:03$ UT (rendered with 13 s in the video). The timestamps mentioned on the 131 {\AA} channel are near-cotemporal to the simulation snapshots. The first column with panels (a), (c), (e), (g), and (i) represents the evolution of field lines neighboring the jet and ribbons whereas the second column with panels (b), (d), (f), (h), and (j) are snapshots of progress of transients in 131 {\AA} channel. Initially, the flux rope (yellow color) starts to untwist due to the action of Lorentz force. The ambient loops in red color close to the flux rope are seen to rise toward the eastern part of the active region along with the flux rope. A close correspondence between the loop dynamics and 131 channel can be noticed in panel (e) and (f) where the flux rope opens up ejecting the materials outward, similar to 131 channel. Afterward, the nearly potential yellow color loops in panel (g) and (h) show the post-reconnected loops. According to the standard picture of a jet in \citet{2015Natur.523..437S}, there are two types of reconnections takes place in a jetting process, the external reconnection which creates a passage for the flux rope to erupt and the internal reconnection involves the reconnection at its footpoints. Here, the escape of the materials is captured while the internal reconnection not so clearly may be owing to lack of spatial resolution and the location of the flux rope being close to the boundary. Although, we see some kinks in the flux rope, as marked by a circle in panel (f) of Figure \ref{ribbon}. However, it is important to note that the simulation is magnetically driven whereas in reality, the jets could be due to the combined effect of both magnetically and thermal changes in the surrounding.

\begin{figure}
\centering
\includegraphics[width=\textwidth]{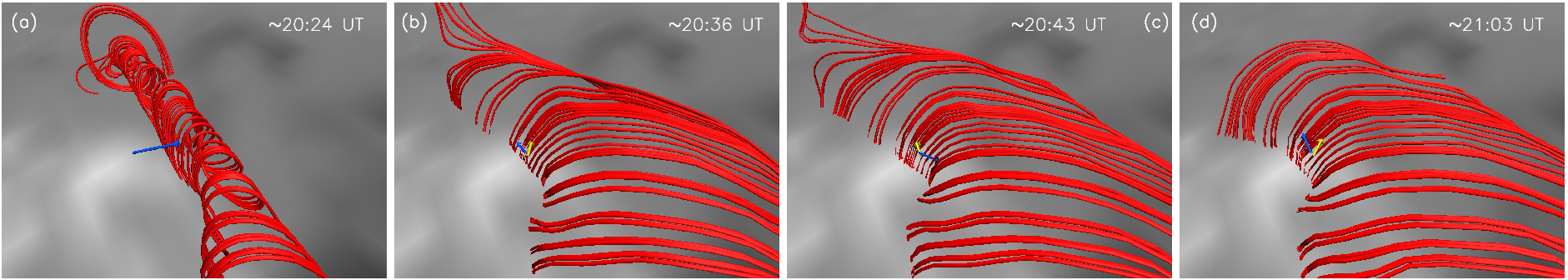}
\caption{Evolution of orientations of Lorentz force (blue arrow) and plasma flow (yellow arrow) to the flux rope (in red color field lines) at a fixed location till the time period coverage as Figure \ref{ribbon}. Initially, the Lorentz force pushes the field lines helping in untwisting it. Important to note is the direction of Lorentz force remaining nearly aligned with the flux rope axis toward the end. Both Lorentz force and flow remain perpendicular to the field line at the end, complementing the profile of angles calculated in Figure \ref{angle}.} 
\label{direction}
\end{figure}

\begin{figure}
\centering
\includegraphics[width=.5\textwidth]{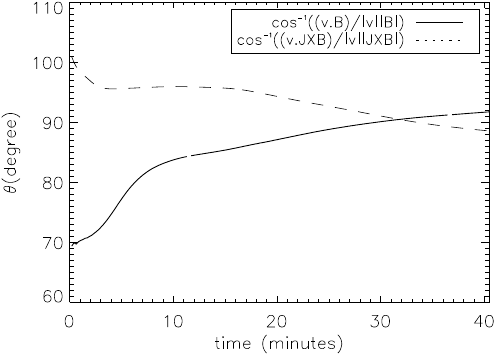}
\caption{Progress of the angle between the flow vector and (1) the magnetic field (solid line), and (2) Lorentz force (dashed line). In both cases, $\theta$ is seen to approach toward $90^{\circ}$ implying the action of plasma flow and Lorentz force on the untwisting of the flux rope field lines. This further supplements the orientation of $\mathbf{v}$, and $\mathbf{J \times B}$ around the flux rope depicted in the second column of Figure \ref{ribbon}.} 
\label{angle}
\end{figure}

Elaborating the discussion on the untwist of the flux rope and the surrounding arcade further, an exclusive evolution of the flux rope is shown in Figure \ref{ribbon} with a comparison of the filament evolution, overplotted with 304 {\AA} channel on the background till the formation of ribbons. The untwisting of the flux rope (pink color) and footpoints of subsequently formed less potential loops which is also evident from the variation of mean of total current density ($|\mathbf{J}|$)  over the simulation period plotted in Figure \ref{current}. They are also seen to follow the ribbons in the later phase of the simulation as shown in the panel (c) of Figure \ref{ribbon}. The directions of Lorentz force and the flow vector are plotted in panels (b), (d), (f) of Figure \ref{ribbon} and denoted in blue and yellow color arrows respectively. Absence of the plasma flow in the first panel is due to the setting to zero in the first time step in the MHD model as it was purely driven by the Lorentz force. Initially downward and inclined Lorentz force pushes the toroidal fluxes leading to untwisting of the flux rope. To focus it more, we investigate the orientation of the Lorentz force and flow vector at a fixed location within the flux rope as shown in Figure \ref{direction}  for different stages of the flux rope evolution. Similar to Figure \ref{ribbon}, the blue arrow depicts the Lorentz force and the pink arrow depicts the flow vector. We appreciate the evolution of the directionality of the Lorentz force (at a fixed spatial location) more clearly as shown in Figure \ref{direction}(a) to (d). This shows that the force remains nearly parallel to the flux rope axis during the untwisting phase of field lines. We also notice that the flow vector remains aligned nearly parallel to the flux rope axis and normal to the magnetic field lines at that fixed location during the untwisting phase of the field lines of the flux rope.  These consistent directionalities during the untwisting phase of the evolution represent that the force and velocity flows in the vicinity of the flux rope play the role in the untwisting of the flux rope field lines. We also estimate the spatial average of twist density ($|T_{w}|$) over a selected region that contains the flux rope as shown in Figure \ref{twist}.  This is also supporting the trend of twist parameter evolution in Figure \ref{twist} in Section \ref{moreen}.
Further supplementing the argument quantitatively, we have plotted the angles between the flow vector ($\bf{v}$),and magnetic field ($\bf{B}$) in solid line, and the Lorentz force ($\bf{J \times B}$) in dashed line respectively in Figure \ref{angle}. The angles are computed as $\cos^{-1}{(\bf{v}\cdot \bf{B})/(|\bf{v})||\bf{B}|)}$ and $\cos^{-1}{(\bf{v}\cdot \bf{J}\times \bf{B})/(|\bf{v})||\bf{J}\times \bf{B}|)}$. In both cases, the angles are approaching 90$^{\circ}$ toward the end of the evolution which is in congruent with the topological evolution seen in Figure \ref{ribbon}. This simultaneously explains the loss of twist and the apparent rise of the field lines in the flux rope as well as the overarching loops there.

\begin{figure}[!h]
\centering
\includegraphics[width=\textwidth]{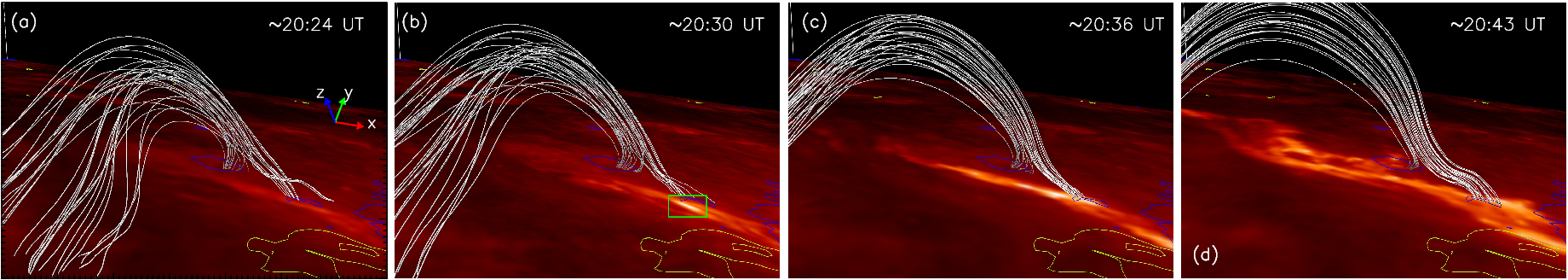}
\includegraphics[width=\textwidth]{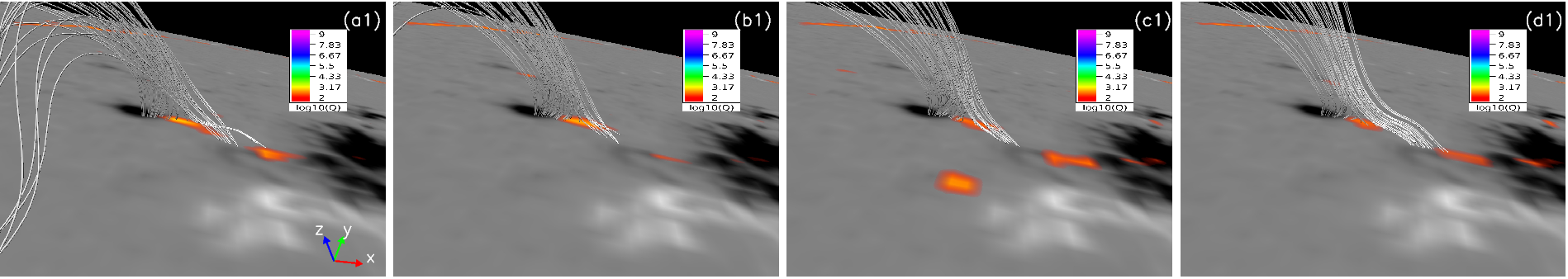}
\caption{Panel (a)-(d): Evolution of the bald patch; field lines colored in white near the bright point structure, and the subsequent flare ribbons. We have overplotted 304 {\AA} channel in the background. The yellow and blue contours represent the positive and negative polarities respectively.  Panel (a1)-(d1): Evolution of $Q$-factor near the bald patch for the same time period. Evident is the reconnection near the bald patch where the field lines are changing the connectivities.}
\label{bpevol}
\end{figure}

In Figure \ref{bpevol}, we have plotted the evolution of the bald patch with 304 {\AA} channel in the background in panels (a)-(d). There, yellow and blue contours in every panel represent positive and negative polarities respectively, and highlight the footpoint connectivities of the loops of the bald patch. In the background, we have plotted the squashing factor or $Q$-value to trace the QSL dynamics in panels (a1)-(d1) for the same time stamps as the first row. In panel (b), we have marked the initial bright point location with the green  color box. The continuous reconnections near this bald patch  along with flux rope eruption in the simulation are suspected to produce the apparent three ribbons later. Also, the footpoint evolutions in a slipping fashion follow the bright point as well as the ribbons as they expand later. Interestingly similar to our case, \citep{2017RAA....17...93J} in their work of a non-typical flare, have found a co-spatiality of flare ribbons and bald patch near the PIL from observations. In their numerical modeling, they found a remarkable agreement between the footpoints of bald patches to the ribbons. In our simulation, albeit the initial set-up of the evolution, the reconnection near the bald patch facilitates the formation of the multiple ribbons which eventually contributed to the jet base region alongside the untwining of flux rope. However, attributing the eruption of this flux rope from the jet base region to the consequent oscillation in the filament on the east side of the active region, as reported in \citep{2023A&A...672A..15J}, is beyond the scope of the simulation presented here. Together, Figure \ref{ribbon} and Figure \ref{bpevol} show the onset of the jet eruption and the formation of the ribbons.

\subsection{More energetics during the evolution near all the ROIs}
\label{moreen}
\begin{figure}[!h]
\centering
\includegraphics[width=.337\textwidth]{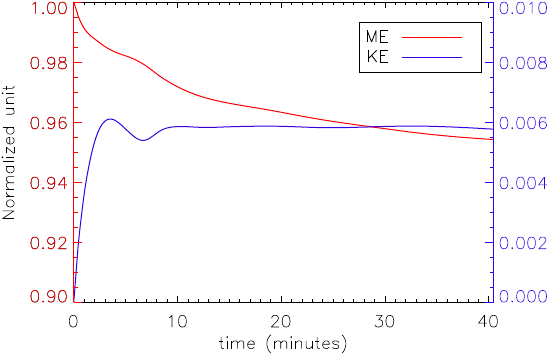}
\includegraphics[width=.325\textwidth]{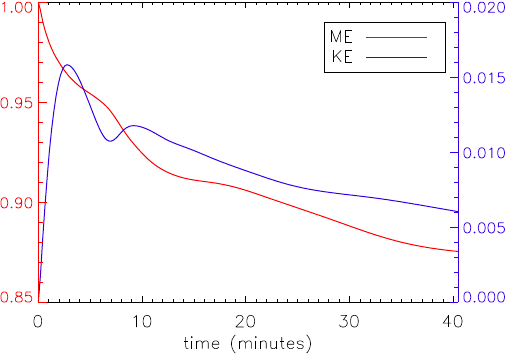}
\includegraphics[width=.325\textwidth]{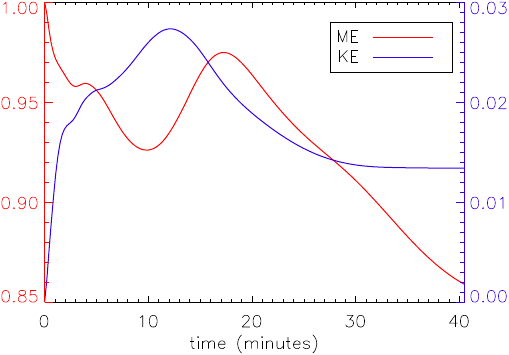}
\caption{Variation of volume averaged magnetic energy density and kinetic energy density normalized over total energy density (a) over the whole computational volume, (b) near the jet base region, and (c) in the neighborhood of the bright point region during the evolution. Evident is the dissipation in the magnetic energy by reconnection while feeding the kinetic energy, during the initial time of the simulation in every panel. Later kinetic energy is also dissipated by the action of viscous relaxation. This feedback relationship is significant in the jet region and more in the bright point region. This suggests a higher rate of dissipation of magnetic energy and conversion to kinetic energy near the locality of the jet base region and the bright point location.}
\label{energy}
\end{figure}

Topological evolution aside, the energetics near the transients are other hints of the energy release mechanism. Hence, in Figure \ref{energy} in panel (a), we have plotted the time variation of volume averaged magnetic energy density (MED; ($B^2/8\pi$)) and kinetic energy density (KED; $v^2/2$,  also $~\rho =1$)  throughout the simulation period for the whole of the computational box. Important is the trend of decrease in the magnetic energy while the kinetic energy is sharply increasing at the expense of it during the first $\approx 4$ minutes of the evolution, decreases between $\approx 5-8$ minutes, and then it increases till 10 minutes. Later, both are evolving toward an almost quasi-steady state. Also, notable is that the initial value of the KED is zero as suggested from the numerical setup of the model in the Section \ref{setup}. In panel (b), we have focused on the base of the jet. The sharp change in the magnetic energy density can be noticed there, $\approx 13\%$ decline to its original strength in the 15 minutes of evolution time scale. However, within the same period of evolution, kinetic energy density is increasing with a rate of $\approx 16\%$ from a motionless state while showing a fall in between $\approx 4-8$ minutes due to viscous dissipation. Further, in the panel (c), we noticed an interesting variation near the neighborhood of the bright point and observed remarkable changes in both MED and KED, where the transients are triggered. The overall decrease in MED during first $\approx 10$ minutes is due to the continuous reconnections near the bald patch. This dissipation may be contributing to the increase in KED.  Noteworthy is the peak in the KED after $\approx 12$ minutes of the start time. The early peak in KED might be responsible for enhancing the twist in the field lines near the bright point as well as in the neighborhood of the flux rope adding more flux to it, which can be seen from panel (b) of Figure \ref{ribbon}. Subsequently, the increasing twist in the field lines assists in increasing the magnetic energy near those locations, which is discernible from the peak of magnetic energy $t\approx 17$ minutes in Figure \ref{energy}(c) after $\approx 5$ minutes to that of kinetic energy. Important is also the continuous reconnection near the bald patch which facilitates the dissipation of magnetic energy along with the triggering of the jet. The KED over every region is dissipated by the action of viscous relaxation. The feedback sharing between the MED and KED  is significant in the jet region and more in the bright point region. This suggests a higher rate of dissipation of magnetic energy and conversion to kinetic energy near the locality of the jet base region and the bright point location. Here, we also draw the attention of the reader toward the efficacy of the ILES scheme of EULAG-MHD model being majorly effective at the locations of primary reconnection sites, whereas not contributing to the other part of the computational box.

\begin{figure}
\centering
\includegraphics[width=.5\textwidth]{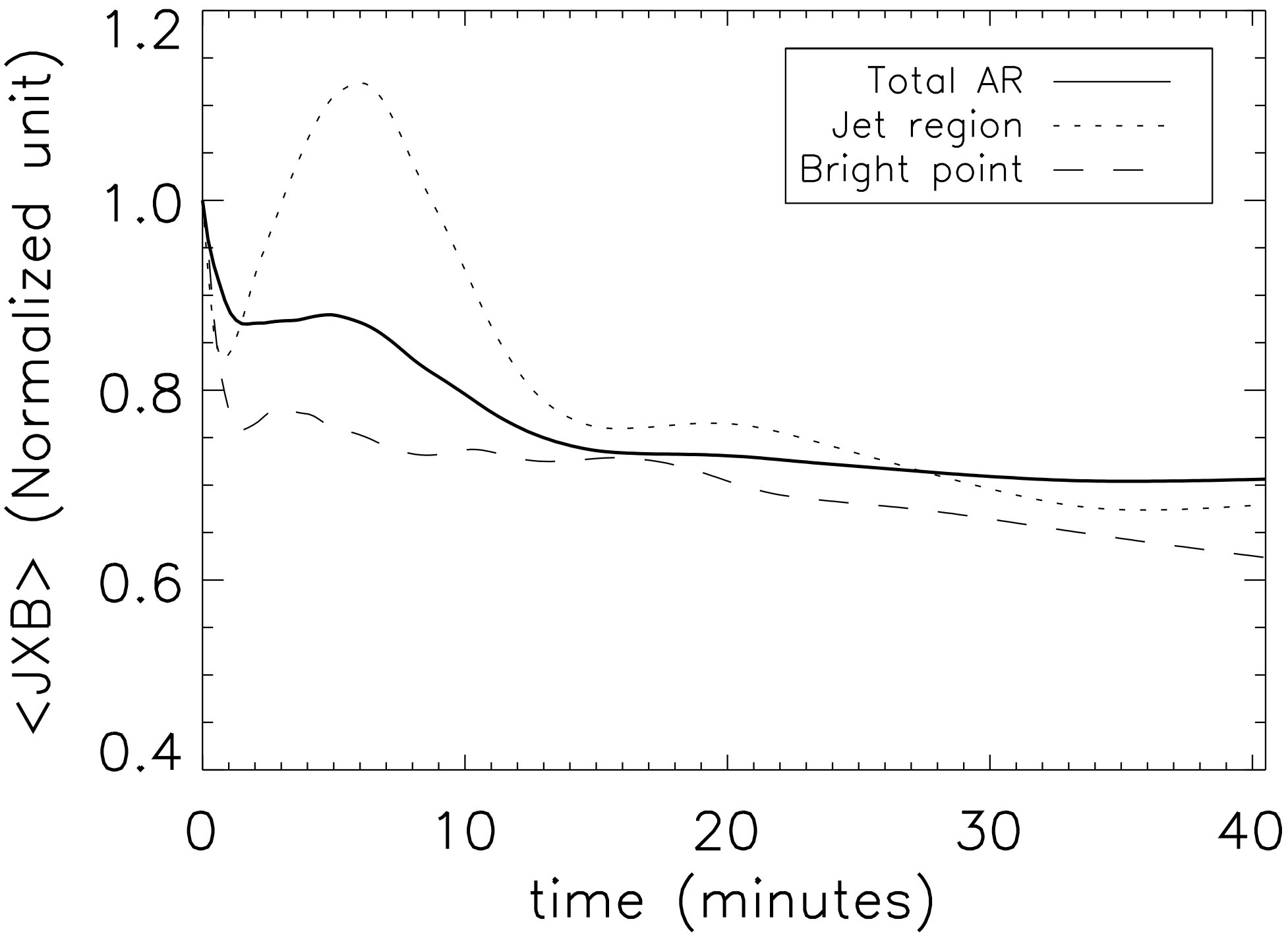}
\caption{Variation of average of Lorentz force normalized to initial value for (a) the whole computational volume in a solid line, (b) the jet base region in dotted lines, and (c) in the neighborhood of the bright point region in dashed lines during the evolution. } 
\label{lforce}
\end{figure}

\begin{figure}
\centering
\includegraphics[width=.5\textwidth]{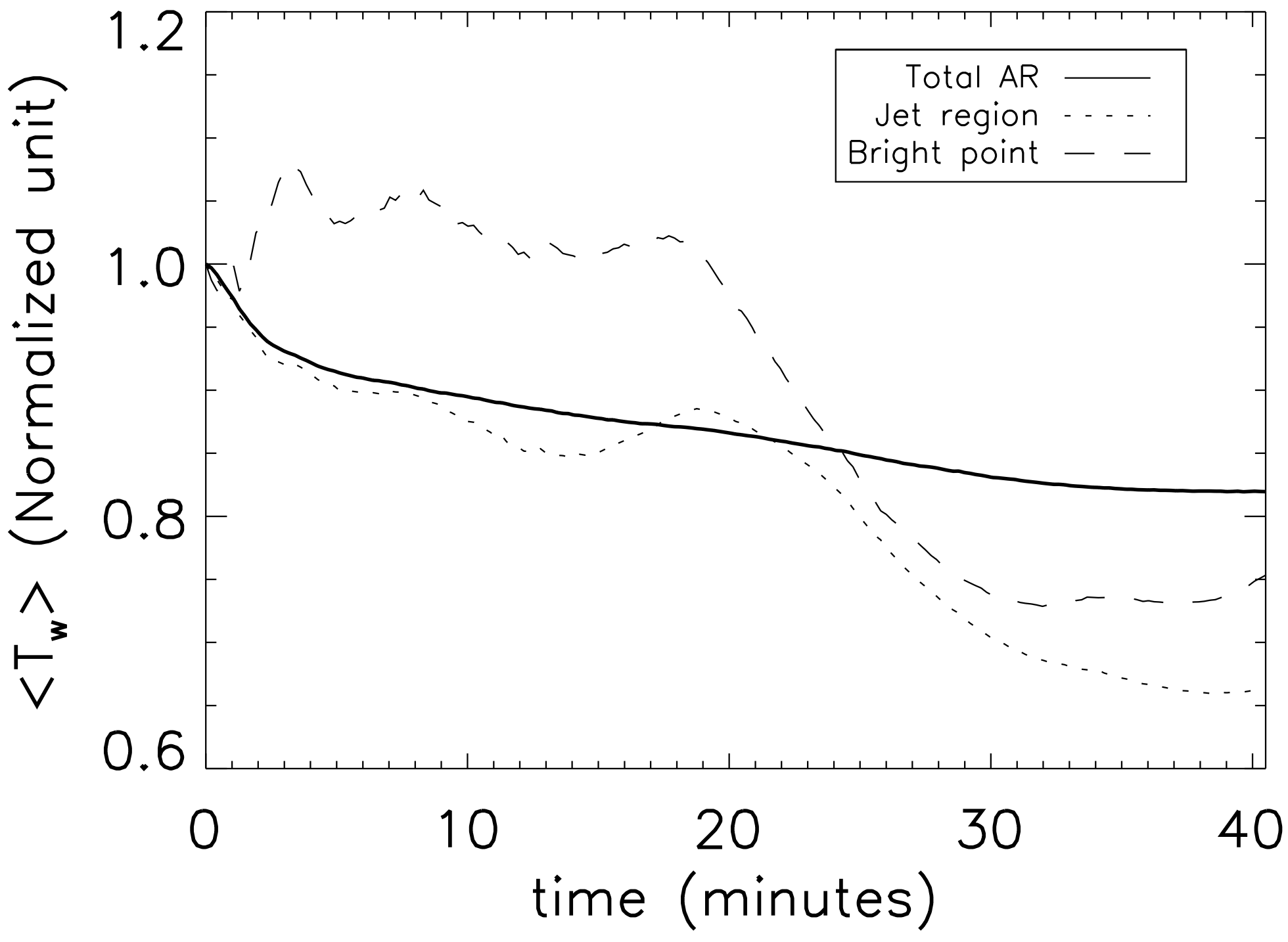}
\caption{Evolution of average of twist density normalized to initial value for (a) the whole computational volume in a solid line, (b) the jet base region in dotted lines, and (c) in the neighborhood of the bright point region in dashed lines during the evolution.} 
\label{twist}
\end{figure}

Next in Figure \ref{lforce}, we have plotted the profile of average of Lorentz force density ($|\mathbf{J\times B}|$) for all ROIs. The profiles are denoted in solid line for the whole of AR, dotted line for the jet region, and the dashed line for the bright point. We observe an initial fall till the first couple of minutes as the $|\mathbf{J\times B}|$ is expended on the generation of the plasma flow which can be seen in the trend of KED in panel (a) of Figure \ref{energy}. A strikingly rise $|\mathbf{J\times B|}$ is seen near the jet region. This might be due to the effect of prior increase of KED which pushes the field lines into the jet region while injecting more twist to flux rope as well as the surrounded arcades. But, it then decreases rapidly due to the unwinding of the flux rope and dissipation near the bright point. The trend of ($|\mathbf{J\times B}|$) is however relatively similar to that of the complete AR till $\approx 17$ minutes the evolution. The continuous reconnection near the bald patch may inhibit the sharp development of Lorentz force near the bright point. Furthermore, in Figure \ref{twist}, we have plotted the variation of mean of twist parameter for the three ROIs. Similar to Figure \ref{lforce}, we have maintained the same legends for three ROIs. The twist in the whole region is decreasing throughout the simulation period whereas, in the case of the jet base region, it increases till $\approx 8$ minutes of the total time, then decreases and maintains almost constancy till the end of the evolution.

\begin{figure}
\centering
\includegraphics[width=.5\textwidth]{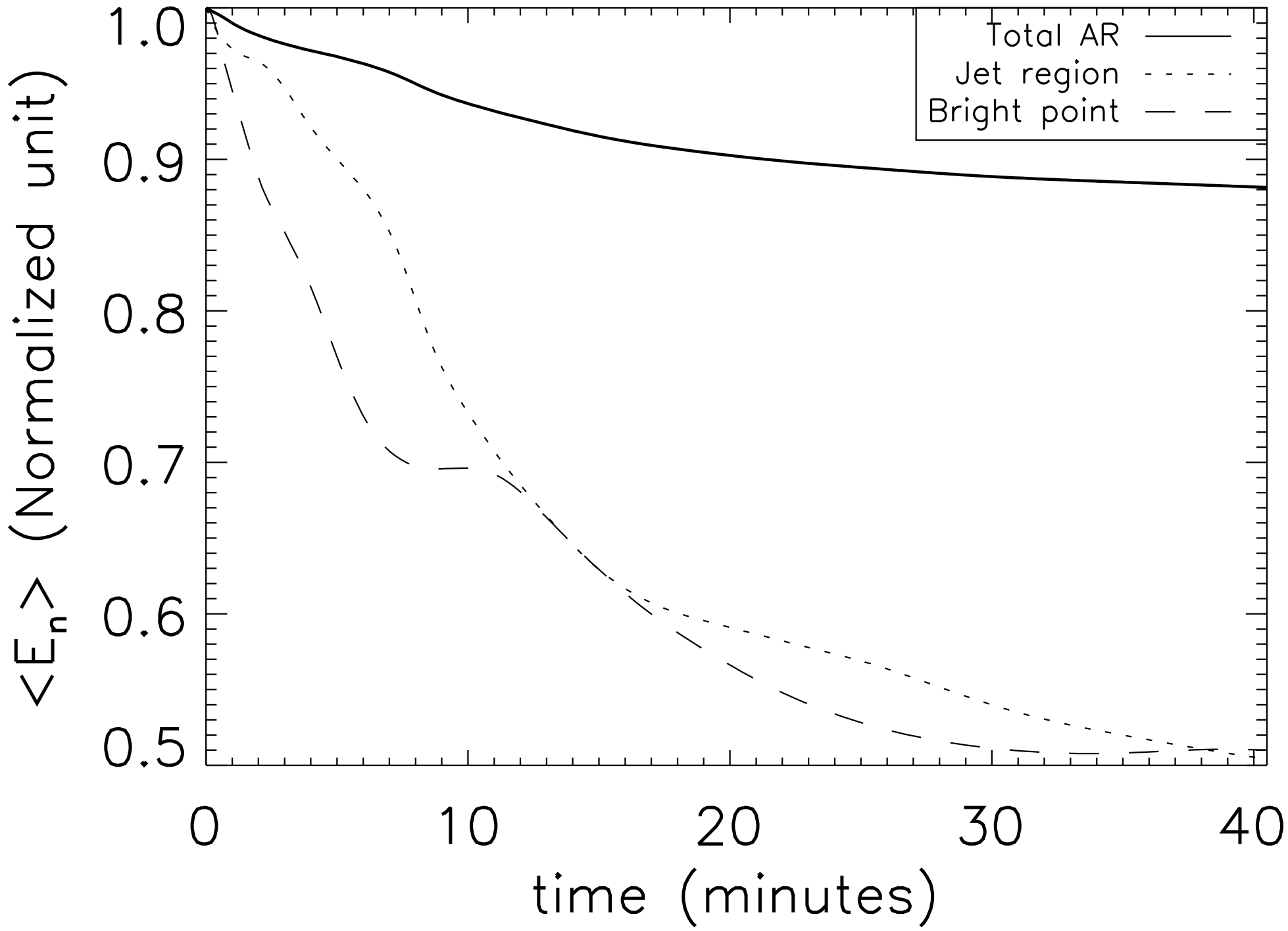}
\caption{Variation of average of free energy normalized to initial value for (a) the whole computational volume in a solid line, (b) the jet base region in dotted lines , and (c) in the neighborhood of the bright point region in dashed lines during the evolution.} 
\label{free_en}
\end{figure}

Next, Figure \ref{free_en} shows the evolution of mean of free energy again for the same ROIs keeping the same legends as the above Figure. We have estimated the free energy in the following way;

\begin{eqnarray}
 E_{free} =  \frac{1}{V}\int_{V} \big \{(B^2/8\pi)_{Sim}\rvert_{t} - (B^2/8\pi)_{Pot}\rvert_{t=0}\} {\rm d}V,  
 \label{free}
\end{eqnarray}
\noindent where, $V$ is the total volume of our simulation box. The first term on the R.H.S of Equation \ref{free} refers to the non-potential energy density at each time step. The second term refers to the potential energy density calculated at the initial time step, as the potential magnetic field is the minimum energy state \citep{1984ApJ...283..349A}. The profiles for the whole AR and jet region are seen to decrease till the end of the evolution. However, for the bright point it shows an increement  between $\approx 7-12$ minutes of the simulation period.

\begin{figure}[ht]
\centering
\includegraphics[width=.325\textwidth]{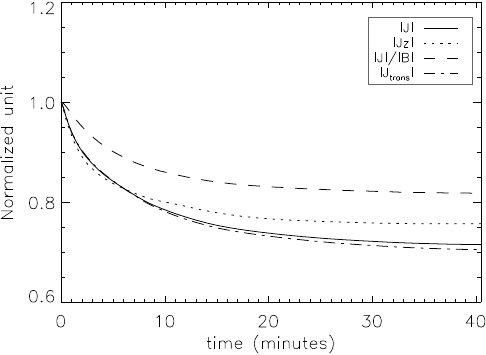}
\includegraphics[width=.31\textwidth]{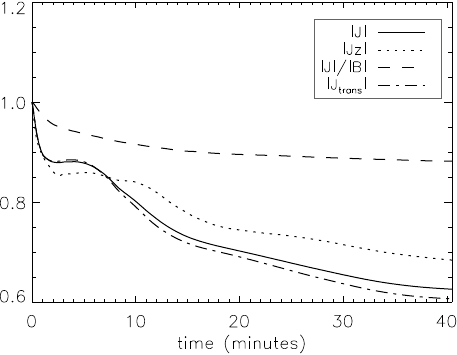}
\includegraphics[width=.31\textwidth]{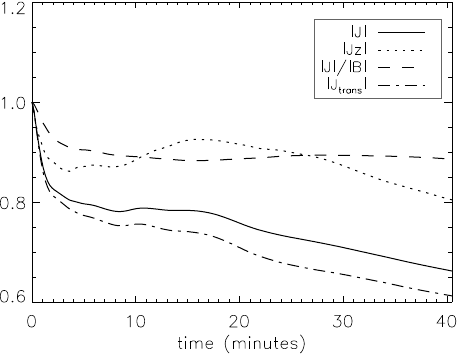}
\caption{Panel (a): the plot shows the variation of volume averaged total current density ($\mathbf{|J|}$), z-component of $\mathbf{|J|}$, $\mathbf{|J|/|B|}$, and $J_{trans}$ or $\sqrt(J_{x}^2+J_{y}^2)$ in solid, dotted, dashed and dash dot lines respectively over the whole computational box. Panel (b): the plot shows the variation of the above parameters near the jet base region. Panel (c): we plot all the above parameters near the bright point. The difference in the variation of $J_{z}$ is noticeable from all the plots, whereas $\mathbf{|J|}$ and $\mathbf{|J|/|B|}$ are displaying an overall decreasing trend throughout the evolution. $J_{trans}$ is apparently following the trend similar to $\mathbf{|J|}$ in every case.} 
\label{current}
\end{figure}

To investigate the energetics further near these locations during the onset and afterward, in Figure \ref{current} we have provided the evolution of volume averaged total current density ($|\mathbf{J}|$), z-component of $\mathbf{J}$, $\mathbf{|J|/|B|}$, and the $J_{trans}$ or $\sqrt{J_{x}^2+J_{y}^2}$ for all the locations stated in the above paragraphs. In panel (a), all the parameters show a monotonous decrease. Near the jet base region and the bright point location, in panel (b) and (c), $J$ and $J_{z}$ show a fall during the initial few minutes, then start picking up due to twisting of the field lines. Interestingly, $J_{z}$ shows an increase in the bright point neighborhood during the peak of the eruption phase which was also the case for magnetic energy density in panel (c) of Figure \ref{energy}. Interestingly, we observe similar growth in the average temperature over the complete activity period in the DEM averaged temperature map, in panel (c) of Figure \ref{demtemp}. This survives till the start of the decay phase of the eruption. After $\approx 30$ minutes of evolution, both are decreasing in a monotonous fashion. However, we fail to capture any abrupt changes in $\mathbf{|J|/|B|}$ during the evolution in any of the regions, particularly near the jet base region. The reason may be partially due to the lower spatial resolutions of computation.  However, locations of current sheet  are not found in the extrapolation resolution scales as well which was of nearly twice of the simulation. Another overt reason can be accredited to the boundary condition employed at the bottom boundary as line-tying.

\section{Summary}
\label{conc}

In this work, we investigate the triggering of an active region jet by magnetic field extrapolation, and its subsequent magnetohydrodynamics evolution through numerical simulation. We have also analyzed the temperature evolution based on the emission measurement from observation near the transients. Then we compare the temporal evolution of temperature (derived from observation) with the heating implication based on MHD simulation. Besides, we have explored the formation of three ribbons on the base of the jet which are atypical.

First, we used the non-force-free-field magnetic field extrapolation to obtain the initial magnetic field topology of the active region. The extrapolated field provides a non-zero Lorentz force in the region at least to a certain height in the lower atmosphere, afterward achieving a nearly force-free state in the upper atmosphere of the computational box. The top boundary of the computational box reaches $\approx 184$ Mm height of the atmosphere. The overall magnetic topology agrees well with the loop morphology from the observation. Near the jet base region, we found a bald patch triggering the jetting process and multiple ribbons. Additionally, a flux rope, though low-lying,  was found in the vicinity of the bright point.

To understand the onset of the jet and the formation of the three ribbons, we utilized the NFFF extrapolated field as an initial state to a data-constrained EULAG-MHD simulation with a line-tying boundary condition on the bottom boundary covering the period of jet eruption. The findings from the simulation are summarized below.
\begin{itemize}
 \item The study provides topological evolution of magnetic field in the neighborhood of the jet, particularly field line evolutions near the flux rope and the bald patch. The untwisting of field lines releases the jet materials along the direction of the plasma flows and ultimately contributes toward the eruption of the jet. The onset of the jet starts due to reconnection in the bald patch which is different from a traditional jet onset mechanism.  
    \item  The simulation simultaneously focuses on the formation of multiple flare ribbons where the same reconnection near the bald patch and the eruption of the flux rope generate multiple ribbons due to the transfer of energetic particles to the lower atmosphere. The footpoint evolution of these topologies are in good agreement with the observed ribbon locations.
    \item Further discussions on the variation of the magnetic and kinetic energies on different parts of the jet not only elucidate the triggering of the jetting process but also validate the credibility of the simulation performed. A detailed focus on the variation of energy densities and the active driving of Lorentz force and modulation by the plasma flow over time near both the jet base and bright point regions explain the relaxation of magnetic energy and kinetic energy in facilitating the triggering and eruption process.  These parametric study highlight the importance of the onset region in order to understand magnetic reconnection. 
    \item Additional highlights are on the profile of the mean of total current and current components which shed light on the current dissipation near those sites. A qualitative comparison with dynamics from the DEM analysis also falls in the line of argument of onset of the jet. We observe a hotter base and cooler spire in the DEM analysis for the jet. The temperature and emission profile near the bright point from the observation, the topological evolution, and the profiles of the simulated energetics comply with the onset process of the jet remarkably. This motivates further to focus on both the magnetic and thermal structure of a jet in order to understand the energy transfer through the jet and as well as in the atmosphere globally.  We noticed a key aspect from the evolution of the current profiles is that the onset point is a maximum current carrying site than the overall transient affected area. However, the observational and/or numerical modeling artifacts are not ignored. Thus, directly quantifying the amount of dissipation into the ohmic heating in these locations may not be so accurate.
	\end{itemize}
Yet, the presented simulation could not (a) capture the formation of the current sheet as similar to the 2D model proposed by \citep{2021ApJ...912...75L}, and as discussed in \citep{2023A&A...672A..15J}. This  may be due to the resolution limit and the boundary conditions adapted in our simulation, and (b) delineate the complete thermodynamics of the jet region due to utilization of an incompressible and thermally inactive model. The jets are ubiquitous and impact not only the local atmosphere but also influence the solar wind generation based on their source of origins. Hence, in our future work, we aim to attempt a more realistic simulation covering all these shortcomings. We will also approach a data-driven boundary condition to the simulation to understand the impact on regularization of both the onset and eruption process of such events, particularly focusing on the genesis of these multiple ribbons.

\vskip 1cm

\facilities{AIA/SDO, HMI/SDO}
\software{Solarsoft \citep{2012:solarsoft}, VAPOR/NCAR}

We thank the anonymous referee for the insightful comments and suggestions which improved the manuscript considerably. The extrapolation and simulations are performed in the Bladerunner cluster located in the Center for Space Plasma and Aeronomic Research department of the University of Alabama in Huntsville. We acknowledge the use of the visualization software VAPOR (www.vapor.ucar.edu) for generating relevant graphics. Data and images are courtesy of NASA/SDO and the HMI and AIA science teams. SDO/HMI is a joint effort of many teams and individuals to whom we are greatly indebted for providing the data. We thank Ranadeep Sarkar for sharing the PIL calculation code. S.S.N. acknowledges NSF-AGS-1954503, NASA-LWS-80NSSC21K0003, and 80NSSC21K1671 grants. S.S. acknowledges support by the European Research Council through the Synergy Grant \#810218 (``The Whole Sun”, ERC-2018-SyG). A.K.S. is supported by funds of the Council of Scientific \& Industrial Research (CSIR), India, under file no. 09/079(2872)/2021-EMR-I. We are grateful to Qiang Hu for his valuable discussions on development of the manuscript.

\bibliography{reference1}

\begin{thebibliography}{}
\expandafter\ifx\csname natexlab\endcsname\relax\def\natexlab#1{#1}\fi
\providecommand{\url}[1]{\href{#1}{#1}}
\providecommand{\dodoi}[1]{doi:~\href{http://doi.org/#1}{\nolinkurl{#1}}}
\providecommand{\doeprint}[1]{\href{http://ascl.net/#1}{\nolinkurl{http://ascl.net/#1}}}
\providecommand{\doarXiv}[1]{\href{https://arxiv.org/abs/#1}{\nolinkurl{https://arxiv.org/abs/#1}}}

\bibitem[{{Altschuler} \& {Newkirk}(1969)}]{1969SoPh....9..131A}
{Altschuler}, M.~D., \& {Newkirk}, G. 1969, \solphys, 9, 131, \dodoi{10.1007/BF00145734}

\bibitem[{{Aly}(1984)}]{1984ApJ...283..349A}
{Aly}, J.~J. 1984, \apj, 283, 349, \dodoi{10.1086/162313}

\bibitem[{{Amari} {et~al.}(1999){Amari}, {Boulmezaoud}, \& {Mikic}}]{1999A&A...350.1051A}
{Amari}, T., {Boulmezaoud}, T.~Z., \& {Mikic}, Z. 1999, \aap, 350, 1051

\bibitem[{{Athiray} \& {Winebarger}(2024)}]{2024ApJ...961..181A}
{Athiray}, P.~S., \& {Winebarger}, A.~R. 2024, \apj, 961, 181, \dodoi{10.3847/1538-4357/ad1837}

\bibitem[{{Aulanier} {et~al.}(2005){Aulanier}, {Pariat}, \& {D{\'e}moulin}}]{2005A&A...444..961A}
{Aulanier}, G., {Pariat}, E., \& {D{\'e}moulin}, P. 2005, \aap, 444, 961, \dodoi{10.1051/0004-6361:20053600}

\bibitem[{{Berger} \& {Prior}(2006)}]{2006JPhA...39.8321B}
{Berger}, M.~A., \& {Prior}, C. 2006, Journal of Physics A Mathematical General, 39, 8321, \dodoi{10.1088/0305-4470/39/26/005}

\bibitem[{{Bhattacharyya} \& {Janaki}(2004)}]{2004PhPl...11.5615B}
{Bhattacharyya}, R., \& {Janaki}, M.~S. 2004, Physics of Plasmas, 11, 5615, \dodoi{10.1063/1.1808454}

\bibitem[{{Bhattacharyya} {et~al.}(2007){Bhattacharyya}, {Janaki}, {Dasgupta}, \& {Zank}}]{2007SoPh..240...63B}
{Bhattacharyya}, R., {Janaki}, M.~S., {Dasgupta}, B., \& {Zank}, G.~P. 2007, \solphys, 240, 63, \dodoi{10.1007/s11207-006-0280-5}

\bibitem[{{Bhattacharyya} {et~al.}(2010){Bhattacharyya}, {Low}, \& {Smolarkiewicz}}]{bhattacharyya+2010phpl}
{Bhattacharyya}, R., {Low}, B.~C., \& {Smolarkiewicz}, P.~K. 2010, Physics of Plasmas, 17, 112901, \dodoi{10.1063/1.3496379}

\bibitem[{{Boerner} {et~al.}(2012){Boerner}, {Edwards}, {Lemen}, {Rausch}, {Schrijver}, {Shine}, {Shing}, {Stern}, {Tarbell}, {Title}, {Wolfson}, {Soufli}, {Spiller}, {Gullikson}, {McKenzie}, {Windt}, {Golub}, {Podgorski}, {Testa}, \& {Weber}}]{2012Boerner}
{Boerner}, P., {Edwards}, C., {Lemen}, J., {et~al.} 2012, \solphys, 275, 41, \dodoi{10.1007/s11207-011-9804-8}

\bibitem[{{Bora} {et~al.}(2022){Bora}, {Bhattacharyya}, {Prasad}, {Joshi}, \& {Hu}}]{2022ApJ...925..197B}
{Bora}, K., {Bhattacharyya}, R., {Prasad}, A., {Joshi}, B., \& {Hu}, Q. 2022, \apj, 925, 197, \dodoi{10.3847/1538-4357/ac3bce}

\bibitem[{{Brueckner} \& {Bartoe}(1983)}]{1983ApJ...272..329B}
{Brueckner}, G.~E., \& {Bartoe}, J. D.~F. 1983, \apj, 272, 329, \dodoi{10.1086/161297}

\bibitem[{{Carmichael}(1964)}]{1964NASSP..50..451C}
{Carmichael}, H. 1964, in NASA Special Publication, Vol.~50, 451

\bibitem[{{Cheung} {et~al.}(2015){Cheung}, {Boerner}, {Schrijver}, {Testa}, {Chen}, {Peter}, \& {Malanushenko}}]{2015Cheung}
{Cheung}, M. C.~M., {Boerner}, P., {Schrijver}, C.~J., {et~al.} 2015, \apj, 807, 143, \dodoi{10.1088/0004-637X/807/2/143}

\bibitem[{{Cheung} \& {DeRosa}(2012)}]{2012ApJ...757..147C}
{Cheung}, M. C.~M., \& {DeRosa}, M.~L. 2012, \apj, 757, 147, \dodoi{10.1088/0004-637X/757/2/147}

\bibitem[{{Chifor} {et~al.}(2008){Chifor}, {Young}, {Isobe}, {Mason}, {Tripathi}, {Hara}, \& {Yokoyama}}]{2008A&A...481L..57C}
{Chifor}, C., {Young}, P.~R., {Isobe}, H., {et~al.} 2008, \aap, 481, L57, \dodoi{10.1051/0004-6361:20079081}

\bibitem[{{Clyne} {et~al.}(2007){Clyne}, {Mininni}, {Norton}, \& {Rast}}]{clyne-2007}
{Clyne}, J., {Mininni}, P., {Norton}, A., \& {Rast}, M. 2007, New Journal of Physics, 9, 301, \dodoi{10.1088/1367-2630/9/8/301}

\bibitem[{{Courant} {et~al.}(1967){Courant}, {Friedrichs}, \& {Lewy}}]{1967IBMJ...11..215C}
{Courant}, R., {Friedrichs}, K., \& {Lewy}, H. 1967, IBM Journal of Research and Development, 11, 215, \dodoi{10.1147/rd.112.0215}

\bibitem[{{Crank} {et~al.}(1947){Crank}, {Nicolson}, \& {Hartree}}]{1947PCPS...43...50C}
{Crank}, J., {Nicolson}, P., \& {Hartree}, D.~R. 1947, Proceedings of the Cambridge Philosophical Society, 43, 50, \dodoi{10.1017/S0305004100023197}

\bibitem[{{Dahlburg} {et~al.}(1991){Dahlburg}, {Antiochos}, \& {Zang}}]{1991ApJ...383..420D}
{Dahlburg}, R.~B., {Antiochos}, S.~K., \& {Zang}, T.~A. 1991, \apj, 383, 420, \dodoi{10.1086/170799}

\bibitem[{{Demoulin} {et~al.}(1996){Demoulin}, {Henoux}, {Priest}, \& {Mandrini}}]{1996A&A...308..643D}
{Demoulin}, P., {Henoux}, J.~C., {Priest}, E.~R., \& {Mandrini}, C.~H. 1996, \aap, 308, 643

\bibitem[{{Freeland} \& {Handy}(2012)}]{2012:solarsoft}
{Freeland}, S.~L., \& {Handy}, B.~N. 2012, {SolarSoft: Programming and data analysis environment for solar physics}, Astrophysics Source Code Library, record ascl:1208.013

\bibitem[{{Gary}(2001)}]{2001SoPh..203...71G}
{Gary}, G.~A. 2001, \solphys, 203, 71, \dodoi{10.1023/A:1012722021820}

\bibitem[{{Gary} \& {Alexander}(1999)}]{1999SoPh..186..123G}
{Gary}, G.~A., \& {Alexander}, D. 1999, \solphys, 186, 123, \dodoi{10.1023/A:1005147921110}

\bibitem[{Grinstein {et~al.}(2007)Grinstein, Margolin, \& Rider}]{grinstein2007book}
Grinstein, F.~F., Margolin, L.~G., \& Rider, W.~J. 2007, Implicit large eddy simulation: computing turbulent fluid dynamics (Cambridge university press), \dodoi{10.1017/CBO9780511618604}

\bibitem[{{Guo} {et~al.}(2016){Guo}, {Xia}, {Keppens}, \& {Valori}}]{2016ApJ...828...82G}
{Guo}, Y., {Xia}, C., {Keppens}, R., \& {Valori}, G. 2016, \apj, 828, 82, \dodoi{10.3847/0004-637X/828/2/82}

\bibitem[{{Hirayama}(1974)}]{1974SoPh...34..323H}
{Hirayama}, T. 1974, \solphys, 34, 323, \dodoi{10.1007/BF00153671}

\bibitem[{{Hong} {et~al.}(2011){Hong}, {Jiang}, {Zheng}, {Yang}, {Bi}, \& {Yang}}]{2011ApJ...738L..20H}
{Hong}, J., {Jiang}, Y., {Zheng}, R., {et~al.} 2011, \apjl, 738, L20, \dodoi{10.1088/2041-8205/738/2/L20}

\bibitem[{{Hu} \& {Dasgupta}(2006)}]{2006GeoRL..3315106H}
{Hu}, Q., \& {Dasgupta}, B. 2006, \grl, 33, L15106, \dodoi{10.1029/2006GL026952}

\bibitem[{{Hu} {et~al.}(2008){Hu}, {Dasgupta}, {Choudhary}, \& {B{\"u}chner}}]{2008ApJ...679..848H}
{Hu}, Q., {Dasgupta}, B., {Choudhary}, D.~P., \& {B{\"u}chner}, J. 2008, \apj, 679, 848, \dodoi{10.1086/587639}

\bibitem[{{Hu} {et~al.}(2010){Hu}, {Dasgupta}, {Derosa}, {B{\"u}chner}, \& {Gary}}]{2010JASTP..72..219H}
{Hu}, Q., {Dasgupta}, B., {Derosa}, M.~L., {B{\"u}chner}, J., \& {Gary}, G.~A. 2010, Journal of Atmospheric and Solar-Terrestrial Physics, 72, 219, \dodoi{10.1016/j.jastp.2009.11.014}

\bibitem[{{Inhester} \& {Wiegelmann}(2006)}]{2006SoPh..235..201I}
{Inhester}, B., \& {Wiegelmann}, T. 2006, \solphys, 235, 201, \dodoi{10.1007/s11207-006-0065-x}

\bibitem[{{Jiang} {et~al.}(2011){Jiang}, {Feng}, {Fan}, \& {Xiang}}]{2011ApJ...727..101J}
{Jiang}, C., {Feng}, X., {Fan}, Y., \& {Xiang}, C. 2011, \apj, 727, 101, \dodoi{10.1088/0004-637X/727/2/101}

\bibitem[{{Jiang} {et~al.}(2012){Jiang}, {Feng}, {Wu}, \& {Hu}}]{2012ApJ...759...85J}
{Jiang}, C., {Feng}, X., {Wu}, S.~T., \& {Hu}, Q. 2012, \apj, 759, 85, \dodoi{10.1088/0004-637X/759/2/85}

\bibitem[{{Jiang} {et~al.}(2017){Jiang}, {Feng}, {Wu}, \& {Hu}}]{2017RAA....17...93J}
{Jiang}, C.-W., {Feng}, X.-S., {Wu}, S.-T., \& {Hu}, Q. 2017, Research in Astronomy and Astrophysics, 17, 093, \dodoi{10.1088/1674-4527/17/9/93}

\bibitem[{{Joshi} {et~al.}(2023){Joshi}, {Luna}, {Schmieder}, {Moreno-Insertis}, \& {Chandra}}]{2023A&A...672A..15J}
{Joshi}, R., {Luna}, M., {Schmieder}, B., {Moreno-Insertis}, F., \& {Chandra}, R. 2023, \aap, 672, A15, \dodoi{10.1051/0004-6361/202245647}

\bibitem[{{Kopp} \& {Pneuman}(1976)}]{1976SoPh...50...85K}
{Kopp}, R.~A., \& {Pneuman}, G.~W. 1976, \solphys, 50, 85, \dodoi{10.1007/BF00206193}

\bibitem[{{Kumar} {et~al.}(2016){Kumar}, {Bhattacharyya}, {Joshi}, \& {Smolarkiewicz}}]{sanjay2016}
{Kumar}, S., {Bhattacharyya}, R., {Joshi}, B., \& {Smolarkiewicz}, P.~K. 2016, \apj, 830, 80, \dodoi{10.3847/0004-637X/830/2/80}

\bibitem[{{Kumar} {et~al.}(2022){Kumar}, {Prasad}, {Sarkar}, \& {Bhattacharyya}}]{2022FrASS...939061K}
{Kumar}, S., {Prasad}, A., {Sarkar}, R., \& {Bhattacharyya}, R. 2022, Frontiers in Astronomy and Space Sciences, 9, 1039061, \dodoi{10.3389/fspas.2022.1039061}

\bibitem[{{Lemen} {et~al.}(2012){Lemen}, {Title}, {Akin}, {Boerner}, {Chou}, {Drake}, {Duncan}, {Edwards}, {Friedlaender}, {Heyman}, {Hurlburt}, {Katz}, {Kushner}, {Levay}, {Lindgren}, {Mathur}, {McFeaters}, {Mitchell}, {Rehse}, {Schrijver}, {Springer}, {Stern}, {Tarbell}, {Wuelser}, {Wolfson}, {Yanari}, {Bookbinder}, {Cheimets}, {Caldwell}, {Deluca}, {Gates}, {Golub}, {Park}, {Podgorski}, {Bush}, {Scherrer}, {Gummin}, {Smith}, {Auker}, {Jerram}, {Pool}, {Soufli}, {Windt}, {Beardsley}, {Clapp}, {Lang}, \& {Waltham}}]{Lemen2012}
{Lemen}, J.~R., {Title}, A.~M., {Akin}, D.~J., {et~al.} 2012, \solphys, 275, 17, \dodoi{10.1007/s11207-011-9776-8}

\bibitem[{{Li} {et~al.}(2019){Li}, {Jaroszynski}, {Pearse}, {Orf}, \& {Clyne}}]{2019Atmos..10..488L}
{Li}, S., {Jaroszynski}, S., {Pearse}, S., {Orf}, L., \& {Clyne}, J. 2019, Atmosphere, 10, 488, \dodoi{10.3390/atmos10090488}

\bibitem[{{Li} {et~al.}(2016){Li}, {Qiu}, {Longcope}, {Ding}, \& {Yang}}]{2016ApJ...823L..13L}
{Li}, Y., {Qiu}, J., {Longcope}, D.~W., {Ding}, M.~D., \& {Yang}, K. 2016, \apjl, 823, L13, \dodoi{10.3847/2041-8205/823/1/L13}

\bibitem[{{Liu} {et~al.}(2016{\natexlab{a}}){Liu}, {Chen}, {Wang}, \& {Liu}}]{2016NatSR...634021L}
{Liu}, R., {Chen}, J., {Wang}, Y., \& {Liu}, K. 2016{\natexlab{a}}, Scientific Reports, 6, 34021, \dodoi{10.1038/srep34021}

\bibitem[{{Liu} {et~al.}(2016{\natexlab{b}}){Liu}, {Kliem}, {Titov}, {Chen}, {Wang}, {Wang}, {Liu}, {Xu}, \& {Wiegelmann}}]{liu-2016}
{Liu}, R., {Kliem}, B., {Titov}, V.~S., {et~al.} 2016{\natexlab{b}}, The Astrophysical Journal, 818, 148, \dodoi{10.3847/0004-637X/818/2/148}

\bibitem[{{Luna} \& {Moreno-Insertis}(2021)}]{2021ApJ...912...75L}
{Luna}, M., \& {Moreno-Insertis}, F. 2021, \apj, 912, 75, \dodoi{10.3847/1538-4357/abec46}

\bibitem[{{MacTaggart} {et~al.}(2021){MacTaggart}, {Prior}, {Raphaldini}, {Romano}, \& {Guglielmino}}]{2021NatCo..12.6621M}
{MacTaggart}, D., {Prior}, C., {Raphaldini}, B., {Romano}, P., \& {Guglielmino}, S.~L. 2021, Nature Communications, 12, 6621, \dodoi{10.1038/s41467-021-26981-7}

\bibitem[{{Mahajan} \& {Yoshida}(1998)}]{1998PhRvL..81.4863M}
{Mahajan}, S.~M., \& {Yoshida}, Z. 1998, Physical Review Letters, 81, 4863, \dodoi{10.1103/PhysRevLett.81.4863}

\bibitem[{{Margolin} {et~al.}(2006){Margolin}, {Rider}, \& {Grinstein}}]{2006JTurb...7...15M}
{Margolin}, L.~G., {Rider}, W.~J., \& {Grinstein}, F.~F. 2006, Journal of Turbulence, 7, 15, \dodoi{10.1080/14685240500331595}

\bibitem[{{Masson} {et~al.}(2009){Masson}, {Pariat}, {Aulanier}, \& {Schrijver}}]{2009ApJ...700..559M}
{Masson}, S., {Pariat}, E., {Aulanier}, G., \& {Schrijver}, C.~J. 2009, \apj, 700, 559, \dodoi{10.1088/0004-637X/700/1/559}

\bibitem[{{McGlasson} {et~al.}(2019){McGlasson}, {Panesar}, {Sterling}, \& {Moore}}]{2019ApJ...882...16M}
{McGlasson}, R.~A., {Panesar}, N.~K., {Sterling}, A.~C., \& {Moore}, R.~L. 2019, \apj, 882, 16, \dodoi{10.3847/1538-4357/ab2fe3}

\bibitem[{{Metcalf} {et~al.}(1995){Metcalf}, {Jiao}, {McClymont}, {Canfield}, \& {Uitenbroek}}]{1995ApJ...439..474M}
{Metcalf}, T.~R., {Jiao}, L., {McClymont}, A.~N., {Canfield}, R.~C., \& {Uitenbroek}, H. 1995, \apj, 439, 474, \dodoi{10.1086/175188}

\bibitem[{{Mitra} {et~al.}(2018){Mitra}, {Joshi}, {Prasad}, {Veronig}, \& {Bhattacharyya}}]{2018ApJ...869...69M}
{Mitra}, P.~K., {Joshi}, B., {Prasad}, A., {Veronig}, A.~M., \& {Bhattacharyya}, R. 2018, \apj, 869, 69, \dodoi{10.3847/1538-4357/aaed26}

\bibitem[{{Mitra} {et~al.}(2023){Mitra}, {Veronig}, \& {Joshi}}]{2023A&A...674A.154M}
{Mitra}, P.~K., {Veronig}, A.~M., \& {Joshi}, B. 2023, \aap, 674, A154, \dodoi{10.1051/0004-6361/202346103}

\bibitem[{{Moon} {et~al.}(2002){Moon}, {Choe}, {Yun}, {Park}, \& {Mickey}}]{2002ApJ...568..422M}
{Moon}, Y.~J., {Choe}, G.~S., {Yun}, H.~S., {Park}, Y.~D., \& {Mickey}, D.~L. 2002, \apj, 568, 422, \dodoi{10.1086/338891}

\bibitem[{{Moore} {et~al.}(2010){Moore}, {Cirtain}, {Sterling}, \& {Falconer}}]{2010ApJ...720..757M}
{Moore}, R.~L., {Cirtain}, J.~W., {Sterling}, A.~C., \& {Falconer}, D.~A. 2010, \apj, 720, 757, \dodoi{10.1088/0004-637X/720/1/757}

\bibitem[{{Moore} {et~al.}(2013){Moore}, {Sterling}, {Falconer}, \& {Robe}}]{2013ApJ...769..134M}
{Moore}, R.~L., {Sterling}, A.~C., {Falconer}, D.~A., \& {Robe}, D. 2013, \apj, 769, 134, \dodoi{10.1088/0004-637X/769/2/134}

\bibitem[{{Moreno-Insertis} {et~al.}(2008){Moreno-Insertis}, {Galsgaard}, \& {Ugarte-Urra}}]{2008ApJ...673L.211M}
{Moreno-Insertis}, F., {Galsgaard}, K., \& {Ugarte-Urra}, I. 2008, \apjl, 673, L211, \dodoi{10.1086/527560}

\bibitem[{{Mulay} {et~al.}(2017){Mulay}, {Del Zanna}, \& {Mason}}]{2017Mulay}
{Mulay}, S.~M., {Del Zanna}, G., \& {Mason}, H. 2017, \aap, 606, A4, \dodoi{10.1051/0004-6361/201730429}

\bibitem[{{Mulay} {et~al.}(2019){Mulay}, {Sharma}, {Valori}, {V{\'a}squez}, {Del Zanna}, {Mason}, \& {Oberoi}}]{2019Mulay}
{Mulay}, S.~M., {Sharma}, R., {Valori}, G., {et~al.} 2019, \aap, 632, A108, \dodoi{10.1051/0004-6361/201936369}

\bibitem[{{Mulay} {et~al.}(2016){Mulay}, {Tripathi}, {Del Zanna}, \& {Mason}}]{2016Mulay}
{Mulay}, S.~M., {Tripathi}, D., {Del Zanna}, G., \& {Mason}, H. 2016, \aap, 589, A79, \dodoi{10.1051/0004-6361/201527473}

\bibitem[{{Nakagawa} {et~al.}(1971){Nakagawa}, {Raadu}, {Billings}, \& {McNamara}}]{1971SoPh...19...72N}
{Nakagawa}, Y., {Raadu}, M.~A., {Billings}, D.~E., \& {McNamara}, D. 1971, \solphys, 19, 72, \dodoi{10.1007/BF00148825}

\bibitem[{{Nayak} {et~al.}(2021){Nayak}, {Bhattacharyya}, \& {Kumar}}]{2021PhPl...28b4502N}
{Nayak}, S.~S., {Bhattacharyya}, R., \& {Kumar}, S. 2021, Physics of Plasmas, 28, 024502, \dodoi{10.1063/5.0035086}

\bibitem[{{Nayak} {et~al.}(2019){Nayak}, {Bhattacharyya}, {Prasad}, {Hu}, {Kumar}, \& {Joshi}}]{2019ApJ...875...10N}
{Nayak}, S.~S., {Bhattacharyya}, R., {Prasad}, A., {et~al.} 2019, \apj, 875, 10, \dodoi{10.3847/1538-4357/ab0a0b}

\bibitem[{{O'Dwyer} {et~al.}(2010){O'Dwyer}, {Del Zanna}, {Mason}, {Weber}, \& {Tripathi}}]{2010ODwyer}
{O'Dwyer}, B., {Del Zanna}, G., {Mason}, H.~E., {Weber}, M.~A., \& {Tripathi}, D. 2010, \aap, 521, A21, \dodoi{10.1051/0004-6361/201014872}

\bibitem[{{Panesar} {et~al.}(2016){Panesar}, {Sterling}, \& {Moore}}]{2016ApJ...822L..23P}
{Panesar}, N.~K., {Sterling}, A.~C., \& {Moore}, R.~L. 2016, \apjl, 822, L23, \dodoi{10.3847/2041-8205/822/2/L23}

\bibitem[{{Panesar} {et~al.}(2017){Panesar}, {Sterling}, \& {Moore}}]{2017ApJ...844..131P}
---. 2017, \apj, 844, 131, \dodoi{10.3847/1538-4357/aa7b77}

\bibitem[{{Panesar} {et~al.}(2018){Panesar}, {Sterling}, \& {Moore}}]{2018ApJ...853..189P}
---. 2018, \apj, 853, 189, \dodoi{10.3847/1538-4357/aaa3e9}

\bibitem[{{Panesar} {et~al.}(2019){Panesar}, {Sterling}, {Moore}, {Winebarger}, {Tiwari}, {Savage}, {Golub}, {Rachmeler}, {Kobayashi}, {Brooks}, {Cirtain}, {De Pontieu}, {McKenzie}, {Morton}, {Peter}, {Testa}, {Walsh}, \& {Warren}}]{2019:panesar}
{Panesar}, N.~K., {Sterling}, A.~C., {Moore}, R.~L., {et~al.} 2019, \apjl, 887, L8, \dodoi{10.3847/2041-8213/ab594a}

\bibitem[{{Pesnell} {et~al.}(2012){Pesnell}, {Thompson}, \& {Chamberlin}}]{pesnell+2012soph}
{Pesnell}, W.~D., {Thompson}, B.~J., \& {Chamberlin}, P.~C. 2012, \solphys, 275, 3, \dodoi{10.1007/s11207-011-9841-3}

\bibitem[{{Prasad} {et~al.}(2017){Prasad}, 2023{Bhattacharyya}, \& {Kumar}}]{prasad+2017apj}
{Prasad}, A., 2023{Bhattacharyya}, R., \& {Kumar}, S. 2017, \apj, 840, 37, \dodoi{10.3847/1538-4357/aa6c58}

\bibitem[{{Prasad} {et~al.}(2018){Prasad}, {Bhattacharyya}, {Hu}, {Kumar}, \& {Nayak}}]{prasad+2018apj}
{Prasad}, A., {Bhattacharyya}, R., {Hu}, Q., {Kumar}, S., \& {Nayak}, S.~S. 2018, \apj, 860, 96, \dodoi{10.3847/1538-4357/aac265}

\bibitem[{{Priest} \& {D{\'e}moulin}(1995)}]{1995JGR...10023443P}
{Priest}, E.~R., \& {D{\'e}moulin}, P. 1995, \jgr, 100, 23443, \dodoi{10.1029/95JA02740}

\bibitem[{{Pucci} {et~al.}(2013){Pucci}, {Poletto}, {Sterling}, \& {Romoli}}]{2013ApJ...776...16P}
{Pucci}, S., {Poletto}, G., {Sterling}, A.~C., \& {Romoli}, M. 2013, \apj, 776, 16, \dodoi{10.1088/0004-637X/776/1/16}

\bibitem[{{Raouafi} {et~al.}(2016){Raouafi}, {Patsourakos}, {Pariat}, {Young}, {Sterling}, {Savcheva}, {Shimojo}, {Moreno-Insertis}, {DeVore}, {Archontis}, {T{\"o}r{\"o}k}, {Mason}, {Curdt}, {Meyer}, {Dalmasse}, \& {Matsui}}]{2016SSRv..201....1R}
{Raouafi}, N.~E., {Patsourakos}, S., {Pariat}, E., {et~al.} 2016, \ssr, 201, 1, \dodoi{10.1007/s11214-016-0260-5}

\bibitem[{{R{\'e}gnier}(2007)}]{2007MmSAI..78..126R}
{R{\'e}gnier}, S. 2007, \memsai, 78, 126

\bibitem[{{R{\'e}gnier}(2013)}]{2013SoPh..288..481R}
---. 2013, \solphys, 288, 481, \dodoi{10.1007/s11207-013-0367-8}

\bibitem[{{Riley} {et~al.}(2006){Riley}, {Linker}, {Miki{\'c}}, {Lionello}, {Ledvina}, \& {Luhmann}}]{2006ApJ...653.1510R}
{Riley}, P., {Linker}, J.~A., {Miki{\'c}}, Z., {et~al.} 2006, \apj, 653, 1510, \dodoi{10.1086/508565}

\bibitem[{{Ruan} {et~al.}(2008){Ruan}, {Wiegelmann}, {Inhester}, {Neukirch}, {Solanki}, \& {Feng}}]{2008A&A...481..827R}
{Ruan}, P., {Wiegelmann}, T., {Inhester}, B., {et~al.} 2008, \aap, 481, 827, \dodoi{10.1051/0004-6361:20078834}

\bibitem[{{Samanta} {et~al.}(2019){Samanta}, {Tian}, {Yurchyshyn}, {Peter}, {Cao}, {Sterling}, {Erd{\'e}lyi}, {Ahn}, {Feng}, {Utz}, {Banerjee}, \& {Chen}}]{2019Sci...366..890S}
{Samanta}, T., {Tian}, H., {Yurchyshyn}, V., {et~al.} 2019, Science, 366, 890, \dodoi{10.1126/science.aaw2796}

\bibitem[{{Sarkar} \& {Srivastava}(2018)}]{2018SoPh..293...16S}
{Sarkar}, R., \& {Srivastava}, N. 2018, \solphys, 293, 16, \dodoi{10.1007/s11207-017-1235-8}

\bibitem[{{Savcheva} {et~al.}(2012){Savcheva}, {Pariat}, {van Ballegooijen}, {Aulanier}, \& {DeLuca}}]{2012ApJ...750...15S}
{Savcheva}, A., {Pariat}, E., {van Ballegooijen}, A., {Aulanier}, G., \& {DeLuca}, E. 2012, \apj, 750, 15, \dodoi{10.1088/0004-637X/750/1/15}

\bibitem[{{Schatten} {et~al.}(1969){Schatten}, {Wilcox}, \& {Ness}}]{1969SoPh....6..442S}
{Schatten}, K.~H., {Wilcox}, J.~M., \& {Ness}, N.~F. 1969, \solphys, 6, 442, \dodoi{10.1007/BF00146478}

\bibitem[{{Schou} {et~al.}(2012){Schou}, {Scherrer}, {Bush}, {Wachter}, {Couvidat}, {Rabello-Soares}, {Bogart}, {Hoeksema}, {Liu}, {Duvall}, {Akin}, {Allard}, {Miles}, {Rairden}, {Shine}, {Tarbell}, {Title}, {Wolfson}, {Elmore}, {Norton}, \& {Tomczyk}}]{2012SoPh..275..229S}
{Schou}, J., {Scherrer}, P.~H., {Bush}, R.~I., {et~al.} 2012, \solphys, 275, 229, \dodoi{10.1007/s11207-011-9842-2}

\bibitem[{{Shen} {et~al.}(2012){Shen}, {Liu}, {Su}, \& {Deng}}]{2012ApJ...745..164S}
{Shen}, Y., {Liu}, Y., {Su}, J., \& {Deng}, Y. 2012, \apj, 745, 164, \dodoi{10.1088/0004-637X/745/2/164}

\bibitem[{{Shibata} {et~al.}(1992){Shibata}, {Ishido}, {Acton}, {Strong}, {Hirayama}, {Uchida}, {McAllister}, {Matsumoto}, {Tsuneta}, {Shimizu}, {Hara}, {Sakurai}, {Ichimoto}, {Nishino}, \& {Ogawara}}]{1992PASJ...44L.173S}
{Shibata}, K., {Ishido}, Y., {Acton}, L.~W., {et~al.} 1992, \pasj, 44, L173

\bibitem[{{Shibata} {et~al.}(2007){Shibata}, {Nakamura}, {Matsumoto}, {Otsuji}, {Okamoto}, {Nishizuka}, {Kawate}, {Watanabe}, {Nagata}, {UeNo}, {Kitai}, {Nozawa}, {Tsuneta}, {Suematsu}, {Ichimoto}, {Shimizu}, {Katsukawa}, {Tarbell}, {Berger}, {Lites}, {Shine}, \& {Title}}]{2007Sci...318.1591S}
{Shibata}, K., {Nakamura}, T., {Matsumoto}, T., {et~al.} 2007, Science, 318, 1591, \dodoi{10.1126/science.1146708}

\bibitem[{{Smolarkiewicz}(2006)}]{smolarkiewicz2006ijnmf}
{Smolarkiewicz}, P.~K. 2006, International Journal for Numerical Methods in Fluids, 50, 1123, \dodoi{10.1002/fld.1071}

\bibitem[{{Smolarkiewicz} \& {Charbonneau}(2013)}]{smolarkiewicz&charbonneau2013jcoph}
{Smolarkiewicz}, P.~K., \& {Charbonneau}, P. 2013, Journal of Computational Physics, 236, 608, \dodoi{10.1016/j.jcp.2012.11.008}

\bibitem[{{Sterling} {et~al.}(2015){Sterling}, {Moore}, {Falconer}, \& {Adams}}]{2015Natur.523..437S}
{Sterling}, A.~C., {Moore}, R.~L., {Falconer}, D.~A., \& {Adams}, M. 2015, \nat, 523, 437, \dodoi{10.1038/nature14556}

\bibitem[{{Sterling} {et~al.}(2017){Sterling}, {Moore}, {Falconer}, {Panesar}, \& {Martinez}}]{2017ApJ...844...28S}
{Sterling}, A.~C., {Moore}, R.~L., {Falconer}, D.~A., {Panesar}, N.~K., \& {Martinez}, F. 2017, \apj, 844, 28, \dodoi{10.3847/1538-4357/aa7945}

\bibitem[{{Sturrock}(1966)}]{1966Natur.211..695S}
{Sturrock}, P.~A. 1966, \nat, 211, 695, \dodoi{10.1038/211695a0}

\bibitem[{{Su} {et~al.}(2018){Su}, {Veronig}, {Hannah}, {Cheung}, {Dennis}, {Holman}, {Gan}, \& {Li}}]{2018Su}
{Su}, Y., {Veronig}, A.~M., {Hannah}, I.~G., {et~al.} 2018, \apjl, 856, L17, \dodoi{10.3847/2041-8213/aab436}

\bibitem[{{Titov}(2007)}]{2007ApJ...660..863T}
{Titov}, V.~S. 2007, \apj, 660, 863, \dodoi{10.1086/512671}

\bibitem[{{Titov} {et~al.}(2002){Titov}, {Hornig}, \& {D{\'e}moulin}}]{2002JGRA..107.1164T}
{Titov}, V.~S., {Hornig}, G., \& {D{\'e}moulin}, P. 2002, Journal of Geophysical Research (Space Physics), 107, 1164, \dodoi{10.1029/2001JA000278}

\bibitem[{{Tiwari}(2012)}]{2012ApJ...744...65T}
{Tiwari}, S.~K. 2012, \apj, 744, 65, \dodoi{10.1088/0004-637X/744/1/65}

\bibitem[{{Wang} \& {Liu}(2012)}]{2012ApJ...760..101W}
{Wang}, H., \& {Liu}, C. 2012, \apj, 760, 101, \dodoi{10.1088/0004-637X/760/2/101}

\bibitem[{{Wang} {et~al.}(2014){Wang}, {Liu}, {Deng}, {Zeng}, {Xu}, {Jing}, \& {Cao}}]{2014ApJ...781L..23W}
{Wang}, H., {Liu}, C., {Deng}, N., {et~al.} 2014, \apjl, 781, L23, \dodoi{10.1088/2041-8205/781/1/L23}

\bibitem[{{Wheatland}(2004)}]{2004SoPh..222..247W}
{Wheatland}, M.~S. 2004, \solphys, 222, 247, \dodoi{10.1023/B:SOLA.0000043579.93988.6f}

\bibitem[{{Wheatland}(2007)}]{2007SoPh..245..251W}
---. 2007, \solphys, 245, 251, \dodoi{10.1007/s11207-007-9054-y}

\bibitem[{{Wiegelmann} \& {Inhester}(2003)}]{2003SoPh..214..287W}
{Wiegelmann}, T., \& {Inhester}, B. 2003, \solphys, 214, 287, \dodoi{10.1023/A:1024282131117}

\bibitem[{{Wiegelmann} {et~al.}(2007){Wiegelmann}, {Neukirch}, {Ruan}, \& {Inhester}}]{2007A&A...475..701W}
{Wiegelmann}, T., {Neukirch}, T., {Ruan}, P., \& {Inhester}, B. 2007, \aap, 475, 701, \dodoi{10.1051/0004-6361:20078244}

\bibitem[{{Wiegelmann} {et~al.}(2017){Wiegelmann}, {Petrie}, \& {Riley}}]{2017SSRv..210..249W}
{Wiegelmann}, T., {Petrie}, G. J.~D., \& {Riley}, P. 2017, \ssr, 210, 249, \dodoi{10.1007/s11214-015-0178-3}

\bibitem[{{Wiegelmann} \& {Sakurai}(2012)}]{2012LRSP....9....5W}
{Wiegelmann}, T., \& {Sakurai}, T. 2012, Living Reviews in Solar Physics, 9, 5, \dodoi{10.12942/lrsp-2012-5}

\bibitem[{{Wiegelmann} {et~al.}(2014){Wiegelmann}, {Thalmann}, \& {Solanki}}]{2014A&ARv..22...78W}
{Wiegelmann}, T., {Thalmann}, J.~K., \& {Solanki}, S.~K. 2014, \aapr, 22, 78, \dodoi{10.1007/s00159-014-0078-7}

\bibitem[{{Yang} {et~al.}(2023){Yang}, {Yan}, {Xue}, {Chen}, {Wang}, {Xu}, \& {Li}}]{2023Yang}
{Yang}, L., {Yan}, X., {Xue}, Z., {et~al.} 2023, \apj, 945, 96, \dodoi{10.3847/1538-4357/acb6f6}

\bibitem[{{Young} \& {Muglach}(2014)}]{2014PASJ...66S..12Y}
{Young}, P.~R., \& {Muglach}, K. 2014, \pasj, 66, S12, \dodoi{10.1093/pasj/psu088}

\bibitem[{{Zhang} {et~al.}(2023){Zhang}, {Musset}, {Glesener}, {Panesar}, \& {Fleishman}}]{2023Zhang}
{Zhang}, Y., {Musset}, S., {Glesener}, L., {Panesar}, N.~K., \& {Fleishman}, G.~D. 2023, \apj, 943, 180, \dodoi{10.3847/1538-4357/aca654}

\bibitem[{{Zhao} \& {Hoeksema}(1993)}]{1993SoPh..143...41Z}
{Zhao}, X., \& {Hoeksema}, J.~T. 1993, \solphys, 143, 41, \dodoi{10.1007/BF00619095}

\bibitem[{{Zhao} \& {Hoeksema}(1994)}]{1994SoPh..151...91Z}
---. 1994, \solphys, 151, 91, \dodoi{10.1007/BF00654084}

\end{thebibliography}

\end{document}